\documentclass[11pt,a4paper]{article}
\usepackage[margin=1.2in]{geometry}

\usepackage{amsmath,amssymb,amsfonts}
\usepackage{graphicx}
\usepackage{placeins}
\usepackage{booktabs}
\usepackage{xcolor}
\usepackage{bm}
\usepackage{slashed}
\usepackage{enumitem}

\newcommand{\TeV}{\ensuremath{\,\text{TeV}}}
\newcommand{\GeV}{\ensuremath{\,\text{GeV}}}

\newcommand{\MB}{{\mathcal{M}^{\mathrm{B}}}}

\newcommand{\Bmat}{\mathcal{B}}
\newcommand{\Cmat}{\mathcal{C}}
\DeclareMathOperator{\tr}{Tr}

\newcommand{\as}{\alpha_s}
\newcommand{\order}[1]{\mathcal{O}(#1)}
\newcommand{\dd}{\mathrm{d}}

\newcommand{\pT}{p_{\mathrm{T}}}

\usepackage{lineno}
\makeatletter
\newcommand*\patchAmsMathEnvironmentForLineno[1]{
  \expandafter\let\csname old#1\expandafter\endcsname\csname #1\endcsname
  \expandafter\let\csname oldend#1\expandafter\endcsname\csname end#1\endcsname
  \renewenvironment{#1}
    {\linenomath\csname old#1\endcsname}
    {\csname oldend#1\endcsname\endlinenomath}}
\newcommand*\patchBothAmsMathEnvironmentsForLineno[1]{
  \patchAmsMathEnvironmentForLineno{#1}
  \patchAmsMathEnvironmentForLineno{#1*}}
\AtBeginDocument{
  \patchBothAmsMathEnvironmentsForLineno{equation}
  \patchBothAmsMathEnvironmentsForLineno{align}
  \patchBothAmsMathEnvironmentsForLineno{gather}
  \patchBothAmsMathEnvironmentsForLineno{flalign}
  \patchBothAmsMathEnvironmentsForLineno{alignat}
  \patchBothAmsMathEnvironmentsForLineno{multline}
}
\makeatother

\newcommand{\KEWSL}{K_{\mathrm{EWSL}}}
\newcommand{\KFO}{K_{\mathrm{FO}}}
\newcommand{\KFOLO}{K^{\mathrm{LO}}_{\mathrm{FO}}}

\newcommand{\code}[1]{\texttt{#1}}

\usepackage{xspace}
\usepackage{mathtools}
\usepackage[expansion=false]{microtype}
\usepackage{cite}
\usepackage{hyperref}
\usepackage{cleveref}

\hypersetup{
    colorlinks=true,
    linkcolor=blue,
    citecolor=blue,
    urlcolor=blue
}

\newcommand{\mgamc}{MadGraph5\_aMC@NLO\xspace}

\newcommand{\mcatnlo}{MC@NLO\xspace}

\newcommand{\LOQCD}{\mathrm{LO}_{\mathrm{QCD}}}
\newcommand{\NLOQCD}{\mathrm{NLO}_{\mathrm{QCD}}}
\newcommand{\NLOEW}{\mathrm{NLO}_{\mathrm{EW}}}

\newcommand{\SDKweak}{\mathrm{SDK_{weak}}}
\newcommand{\EWSL}{\mathrm{EWSL}}

\renewcommand{\Re}{\operatorname{Re}}

\providecommand{\beqn}{\begin{eqnarray}}
\providecommand{\eeqn}{\end{eqnarray}}
\providecommand{\beq}{\begin{equation}}
\providecommand{\eeq}{\end{equation}}

\begin{document}

\title{Resonance-aware electroweak Sudakov reweighting for multijet-merged predictions}

\author{Lenart Jerala$^{1, 2}$, Rikkert Frederix$^3$, Miha Mu\v{s}kinja$^{1,2}$,\\
Davide Pagani$^4$, Timea Vitos$^{5,6}$, Marco Zaro$^7$\\[4pt]
\small $^1$Jo\v{z}ef Stefan Institute, Jamova cesta 39, SI-1000 Ljubljana, Slovenia\\
\small $^2$Faculty of Mathematics and Physics, University of Ljubljana,\\
\small Jadranska ulica 19, SI-1000 Ljubljana, Slovenia\\
\small $^3$Department of Physics, Lund University, S\"olvegatan 14A, SE-223 62 Lund, Sweden\\
\small $^4$INFN, Sezione di Bologna, Via Irnerio 46, I-40126 Bologna, Italy\\
\small $^5$Department of Physics and Astronomy, Uppsala University,\\
\small Box 516, SE-751 20 Uppsala, Sweden\\
\small $^6$Institute for Theoretical Physics, ELTE E\"otv\"os Lor\'and University,\\
\small P\'azm\'any P\'eter s\'et\'any 1/A, H-1117 Budapest, Hungary\\
\small $^7$TIFLab, Universit\`a degli Studi di Milano\\
\small and INFN, Sezione di Milano, Via Celoria 16, I-20133 Milano, Italy}

\date{}

\maketitle

\begin{abstract}\noindent
We present a resonance-aware extension of electroweak Sudakov-logarithm
reweighting in \mgamc{} for NLO QCD-matched and FxFx multijet-merged
event samples. The method identifies the hard production process
underlying each event and evaluates its electroweak Sudakov correction
while accounting for the spin correlations between unstable particles
and their decay products.
Photonic logarithms are removed from the virtual weight to avoid overlap
with QED shower evolution.

We validate the method in FxFx-merged $Z+\mathrm{jets}$ production and
fully leptonic $W^+W^-$ production. The reweighting yields the expected
growing weak suppression in high-energy tails and is stable under
variations of the clustering and merging parameters. In $W^+W^-$
production, the density-matrix and spin-summed prescriptions are
statistically compatible for the observables studied, indicating only
a small additional helicity dependence of the Sudakov correction in
this channel. The method provides a practical virtual weak-Sudakov
approximation for matched and merged simulations, but does not
constitute an exact NLO electroweak calculation.
\end{abstract}

\section{Introduction}
\label{sec:introduction}

Fixed-order calculations provide systematically improvable predictions
for partonic observables, but realistic exclusive simulations at the
Large Hadron Collider (LHC) also require the resummation of soft and
collinear radiation, hadronisation, and resonance decays.
Next-to-leading-order (NLO) calculations matched to parton showers (PS)
therefore form the standard baseline, while multijet
merging is needed when several hard parton multiplicities must be
described simultaneously~\cite{Frixione:2002ik,Nason:2004rx,Frixione:2007vw,Alioli:2010xd,Platzer:2011bc,Hamilton:2012np,Hoeche:2012yf,Frederix:2012ps,Alwall:2014hca,Bierlich:2022pfr,Sherpa:2024mfk,Bellm:2025pcw}. In
current precision workflows, quantum chromodynamics (QCD) corrections
are usually treated at NLO+PS or multijet-merged NLO+PS accuracy, while
quantum electrodynamics (QED) radiation is commonly left
to the parton shower. This combination provides an accurate
description of the dominant strong corrections and the associated soft
and collinear radiation pattern, and it is the basis of most realistic
collider simulations.

The situation is less mature on the electroweak side. Exact
fixed-order NLO electroweak corrections are now fully automated for a
broad class of processes~\cite{Hirschi:2011pa,Cullen:2011ac,Cascioli:2011va,Actis:2012qn,Kallweit:2014xda,Frixione:2015zaa,Chiesa:2015mya,Actis:2016mpe,Denner:2017wsf,Biedermann:2017yoi,Chiesa:2017gqx,Frederix:2018nkq,Buccioni:2019sur,Pagani:2021iwa}, but their
matching to parton showers and their extension to merged samples have
been studied only in selected processes and frameworks~\cite{Barze:2012tt,Barze:2013fru,Kallweit:2015dum,Muck:2016pko,Granata:2017iod,Gutschow:2018tuk,Chiesa:2019ulk,Chiesa:2020ttl,Brauer:2020kfv,Lindert:2022qdd,Flower:2026byh,Flower:2026lui}.

At high energies, the dominant virtual electroweak effects are
captured by electroweak Sudakov logarithms (EWSL), which contain
double and single logarithms of hard invariants over the weak
scale~\cite{Denner:2000jv,Denner:2001gw,Denner:2003wi,Denner:2004iz,Denner:2006jr,Denner:2008yn}. These
logarithms provide a practical approximation to the leading
high-energy electroweak corrections, but their applicability requires
all relevant invariants to be large compared with the electroweak
scale. At the LHC, this hierarchy can fail because additional QCD
partons introduce pairwise invariants below the weak scale, or because
resolved decay products reconstruct an unstable resonance. Applying
the logarithmic correction directly to such a final state can
therefore identify the wrong external hard object and cannot, in
general, reproduce the helicity-dependent reweighting of
production-decay correlations.

Automated implementations of electroweak Sudakov logarithms already
exist in several environments. The first was provided in
\textsc{Sherpa}~\cite{Bothmann:2020sxm}. The implementation in
\textsc{MadGraph5\_aMC@NLO}~\cite{Pagani:2021vyk,Pagani:2023wgc}
features several improvements over the original algorithm.
\textsc{OpenLoops} provides implementations at one
loop~\cite{Lindert:2023fcu} and at two loops~\cite{Lindert:2026ief}.

A key starting point for the present work is the electroweak Sudakov
reweighting module of \textsc{MadGraph5\_aMC@NLO}, developed in
ref.~\cite{Pagani:2023wgc}. In that implementation, the logarithmic EW
Sudakov approximation is evaluated event by event on the external legs
of an MC@NLO hard process and used as a multiplicative correction
factor, providing a fast approximation to the dominant high-energy
electroweak effects. Photonic logarithms are removed from the virtual
weight to avoid overlap with QED shower evolution; real QED radiation
is delegated to the shower or to a dedicated fixed-order treatment.
In its original form, however, the reweighting is applied to the
resolved Born-like configuration associated with the hard process, and
therefore does not yet provide a prescription for reconstructing the
Sudakov hard object in events with multiple additional QCD emissions,
nor a treatment of the spin correlations associated with heavy
resonances, such as $W$ and $Z$ bosons and top quarks. These two
limitations are precisely the ones addressed in the present paper.  A
resonance-aware treatment of the spin correlations is available in the
\textsc{OpenLoops} implementation~\cite{Lindert:2023fcu}.  Electroweak
Sudakov corrections have also been combined with multijet merging in
\textsc{Sherpa} for $ZZ$ and $ZZj$
production~\cite{Bothmann:2021led}. In this work, we present
similar efforts within the \textsc{MadGraph5\_aMC@NLO} framework.

In this work we extend the \mgamc{} Sudakov reweighting in
two directions. First, for FxFx-merged samples, we replay the
clustering history assigned during the merging procedure and use it to
reconstruct the Born-like object on which the electroweak Sudakov
invariants are evaluated. This avoids applying the high-energy
logarithmic operator to configurations in which matrix-element QCD
radiation or resolved decay products introduce electroweak-scale
invariants. Second, when this reconstruction exposes an unstable
spin-carrying resonance, such as a $W$, $Z$, or top quark, we
replace the spin-summed Sudakov factor by a helicity-density-matrix
weight. The production matrix is corrected in resonance-helicity space
and then contracted with the decay matrices, so that production-decay
spin correlations already present in the event are retained while
their helicity-dependent electroweak reweighting is included. The
result is a QED-subtracted virtual weak Sudakov correction applied
multiplicatively to NLO QCD-matched or FxFx-merged event samples.

The paper is organised as follows. Section~\ref{sec:framework}
introduces the electroweak Sudakov approximation and presents the
resonance-aware reweighting formalism, including the clustering replay
and the density-matrix treatment of decaying resonances.
Section~\ref{sec:validation} validates the implementation on
representative $Z+\mathrm{jets}$ and $W^+W^-$ samples.
Section~\ref{sec:conclusions} summarises the results and discusses
possible extensions.

\section{Electroweak Sudakov reweighting with resonance-aware FxFx clustering}
\label{sec:ewsl_framework}
\label{sec:framework}
\label{sec:method}

\subsection{Electroweak Sudakov logarithms}
\label{sec:la_structure}
\label{sec:sudakov_domain}
\label{sec:sdkweak}
\label{sec:ewsl_ps}

At energies well above the electroweak scale, virtual electroweak corrections
contain double and single Sudakov logarithms of the schematic form~\cite{Denner:2000jv,Denner:2001gw}
\begin{equation*}
  \frac{\alpha}{4\pi}
  \log^2\!\left(\frac{s}{M_W^2}\right),
  \qquad
  \frac{\alpha}{4\pi}
  \log\!\left(\frac{s}{M_W^2}\right).
\end{equation*}
We denote these logarithmically enhanced terms collectively by
$\EWSL$. Their automation in \mgamc{} has been described in
refs.~\cite{Pagani:2021vyk,Pagani:2023wgc}. Since the approximation is
formulated at amplitude level, it also provides the natural starting
point for the resonance-aware density-matrix construction developed
below.

For an $n$-leg process, the Born amplitude $\MB$ carries the gauge
group indices of all external particles,
\begin{equation}
  \MB{}^{\,i_1\cdots i_n}(p_1,\ldots,p_n).
  \label{eq:la_born}
\end{equation}
At $\order{\alpha}$, the logarithmically enhanced one-loop correction can be
written as
\begin{equation}
  \delta\MB{}^{\,i_1\cdots i_n}(p_1,\ldots,p_n)
  =
  \MB{}^{\,i_1'\cdots i_n'}(p_1,\ldots,p_n)\,
  \delta^{\,i_1'i_1\cdots i_n'i_n},
  \label{eq:la_factorised}
\end{equation}
with an implicit sum over the primed indices. The operator $\delta$
acts in the space of external-particle quantum numbers and is
therefore not, in general, a universal scalar multiplying the original
Born amplitude. In particular, the electroweak generators entering the
logarithmic correction may relate the original process to amplitudes
with different weak-isospin components. A scalar correction is
obtained only after the interference with the appropriate Born
amplitudes.

The one-loop logarithmic operator is conventionally decomposed as
\begin{equation}
  \delta
  =
  \delta_{\mathrm{LSC}}
  +
  \delta_{\mathrm{SSC}}
  +
  \delta_{\mathrm C}
  +
  \delta_{\mathrm{PR}}.
  \label{eq:la_decomp}
\end{equation}
The LSC term contains the angular-independent leading soft-collinear
contributions, including the Sudakov double logarithms, while the SSC
term contains angular-dependent subleading soft-collinear single
logarithms. The C term collects single logarithms of collinear origin,
including external-leg field-renormalisation and, where relevant,
Yukawa-enhanced terms.
The PR term contains logarithmic contributions from the renormalisation
of electroweak parameters, including couplings, masses, and mixing
parameters.

The external object on which $\delta$ acts fixes the kinematic
arguments of the logarithms. In the all-incoming convention used in
the EWSL
formulation~\cite{Denner:2000jv,Denner:2001gw,Pagani:2021vyk,Pagani:2023wgc},
the relevant two-leg invariants are
\begin{equation}
  r_{ij}
  \equiv
  \left(p_i+p_j\right)^2
  =
  m_i^2+m_j^2+2\,p_i\!\cdot\!p_j,
  \label{eq:rij}
\end{equation}
for the pairs of external legs that enter the logarithmic kernels. Crossing a
physical $2\to n$ process into this convention reverses the momenta of the
outgoing legs, so $r_{ij}$ is negative for $t$- and $u$-channel-like
pairings. The real parts of the logarithms therefore involve
$\lvert r_{ij}\rvert$, while the sign of the crossed invariant is retained
where it determines the analytic-continuation phase.

For a phase-space configuration $\Phi$, let $\mathcal K(\Phi)$ denote the set of
external-leg pairs used by the logarithmic operator. The formal high-energy
hierarchy is
\begin{equation}
  \left|r_{ij}\right|
  \gg
  M_{\mathrm{EW}}^2,
  \qquad
  M_{\mathrm{EW}}^2
  \sim M_W^2\sim M_Z^2\sim M_H^2\sim m_t^2,
  \qquad
  (i,j)\in\mathcal K(\Phi).
  \label{eq:sudakov_domain_strict}
\end{equation}
This hierarchy cannot be imposed strictly throughout a complete
hadron-collider phase space. Following the automated
implementation~\cite{Pagani:2021vyk,Pagani:2023wgc}, we therefore use
the operational inequality
\begin{equation}
  \min_{(i,j)\in\mathcal K(\Phi)}
  \left|r_{ij}\right|
  \ge M_W^2,
  \label{eq:sudakov_domain}
\end{equation}
as a necessary condition for evaluating the logarithmic correction on
a candidate hard configuration. As in the implementation of EWSL
corrections for MC@NLO processes~\cite{Pagani:2023wgc}, this condition
is applied to pre-shower events, possibly clustered to a
lower-multiplicity state, rather than to the fully showered final
state.

The usual spin- and colour-summed logarithmic correction is obtained
from the Born--virtual interference,
\begin{equation}
  \delta^{\mathrm{EW}}_{\mathrm{LA}}
  \equiv
  \frac{
    2\,\Re\!\left(
      \MB^*\,\delta\MB
    \right)
  }{
    \left|\MB\right|^2
  },
  \qquad
  w
  =
  1+\delta^{\mathrm{EW}}_{\mathrm{LA}},
  \label{eq:delta_ew}
\end{equation}
where the sums and averages over all external quantum numbers are
implicit, LA denotes the logarithmic approximation, and $w$ is the
event-reweighting factor. The event-level
construction is multiplicative. For a hard event with kinematics
$\Phi$, let $\widehat\Phi_{\mathrm B}$ be the Born-like object on
which the EWSL correction is evaluated; it will be defined in
\cref{sec:baseline}. We then write schematically
\begin{equation}
  \dd\sigma^{\mathrm{EWSL}}
  (\Phi)
  =
  w\!\left(\widehat\Phi_{\mathrm B}[\Phi]\right)
  \dd\sigma(\Phi),
  \label{eq:multiplicative_weight}
\end{equation}
where $w$ is the reweighting factor in \cref{eq:delta_ew}.

Upon QCD showering, multiplying the hard-event weight by a relative
electroweak correction of order $\alpha$ generates (additional) mixed
terms of the form
\begin{equation*}
  \alpha\,\as^n,
  \qquad n>0,
\end{equation*}
relative to the leading result. These terms lie beyond the fixed-order
logarithmic approximation and form part of the intended multiplicative
estimate of mixed higher-order effects.

When the shower also contains QED radiation, there is a potential
overlap with the virtual photonic contribution. This can be avoided
with the $\SDKweak$ prescription of
refs.~\cite{Pagani:2021vyk,Pagani:2023wgc}. It removes logarithms of
genuinely photonic origin while retaining weak logarithms and the
ultraviolet parameter-renormalisation terms. The resulting prediction
is therefore a QED-subtracted virtual weak Sudakov approximation
applied to an NLO QCD-matched or FxFx-merged sample. It is not an
exact NLO electroweak prediction: it does not include the full
fixed-order QED contribution, photon-induced
channels beyond those already present in the underlying sample, or
non-logarithmic one-loop electroweak terms.

\subsection{FxFx clustering and reconstruction of the Born-like object}
\label{sec:baseline}

The logarithmic operator requires a definite set of external momenta and
quantum numbers. In a fixed-multiplicity calculation this set is supplied by
the corresponding matrix element. In an FxFx-merged sample, however, the
pre-shower Les Houches Event (LHE) record can originate from any of
several matched multiplicities
and can contain several matrix-element partons. The Sudakov correction is
therefore evaluated on a Born-like object reconstructed from the assigned FxFx
history.

FxFx combines independently generated \mcatnlo{}-matched samples for
the process family
\begin{equation}
  \mathcal P_i:
  \qquad
  ab\to S+i\ \text{partons},
  \qquad i=0,\ldots,N,
  \label{eq:fxfx_process_family}
\end{equation}
where $S$ denotes the hard final-state system of interest and contains no
massless QCD partons. If the Born contribution to the
zero-parton process is of order $\as^b$, then the Born contribution to
$\mathcal P_i$ is of order $\as^{b+i}$.

For the \mcatnlo{} sample whose underlying Born process is
$\mathcal P_i$, we write
\begin{equation}
  \dd\sigma^{\mathrm{MC@NLO}}_{[i]}
  =
  \dd\sigma^{\mathbb S}_{[i]}
  +
  \dd\sigma^{\mathbb H}_{[i]}.
  \label{eq:mcatnlo_sample}
\end{equation}
The bracketed index $[i]$ labels the underlying Born multiplicity. The
$\mathbb S$ contribution has Born-like kinematics,
\begin{equation*}
  \mathbb S_{[i]}:
  \qquad
  ab\to S+i\ \text{partons},
\end{equation*}
whereas the $\mathbb H$ contribution contains one additional parton,
\begin{equation*}
  \mathbb H_{[i]}:
  \qquad
  ab\to S+(i+1)\ \text{partons}.
\end{equation*}
After showering, both classes can populate higher jet multiplicities, but their
fixed-order origins remain distinct.

Neighbouring matched samples contain common tree-level configurations. In
particular, the tree-level contribution with $i$ additional QCD partons is
contained both in the Born-like part of the $i$-parton sample and in the
real-emission part of the $(i-1)$-parton sample,
\begin{equation}
  \dd\sigma^{\mathbb S}_{[i]}
  \supset
  \dd\sigma^{(0)}_i,
  \qquad
  \dd\sigma^{\mathbb H}_{[i-1]}
  \supset
  \dd\sigma^{(0)}_i.
  \label{eq:fxfx_tree_overlap}
\end{equation}
Here the superscript $(0)$ denotes the tree-level contribution, while
the subscript $i$ counts the additional QCD partons.
FxFx assigns the corresponding resolved regions of phase space using a
jet-resolution variable and a merging scale, denoted by
$Q_{\mathrm{cut}}$. Multiplicities with $i<N$ are made exclusive with respect
to sufficiently hard additional radiation, while the highest
multiplicity remains inclusive above $Q_{\mathrm{cut}}$ because no
higher-multiplicity matrix-element sample is present. In both cases,
the hard event is clustered step by step to apply the partial QCD
no-emission suppression and $\as$
reweighting~\cite{Frederix:2012ps}. This clustering includes all
particles in the process, not only the QCD
partons~\cite{Frederix:2021agh}, and continues until the core process
is found.

For the electroweak reweighting, we reuse the sequence of clustering
steps assigned by FxFx. Each step combines particles into a candidate
parent and maps their momenta onto an on-shell configuration with
fewer external particles. This sequence, which we refer to as the
clustering history $\mathcal H$, provides the candidate hard
configurations,
\begin{equation}
  \mathcal H:
  \qquad
  \Phi^{(0)}
  \longrightarrow
  \Phi^{(1)}
  \longrightarrow
  \cdots
  \longrightarrow
  \Phi^{(K)}.
  \label{eq:history_sequence}
\end{equation}
That is, the pre-shower configuration is denoted by $\Phi^{(0)}$.  We
then apply the assigned clustering steps in order, obtaining
configurations of progressively lower multiplicity until reaching the
core $\Phi^{(K)}$. At level $k$, let $p_a^{(k)}$ denote the momentum
of the current external object $a$ mapped to its mass shell, and let
$\mathcal K_k\equiv\mathcal K(\Phi^{(k)})$ be the set of pairs
entering the Sudakov operator on that candidate configuration. We
define
\begin{equation}
  Q_{\mathrm{EW}}^2\!\left(\Phi^{(k)}\right)
  \equiv
  \min_{(a,b)\in\mathcal K_k}
  \left|
    r_{ab}^{(k)}
  \right|,
  \qquad
  r_{ab}^{(k)}
  =
  \left(p_a^{(k)}+p_b^{(k)}\right)^2,
  \label{eq:history_hardness}
\end{equation}
using the same all-incoming invariant convention as in \cref{eq:rij}.

We traverse the history $\mathcal H$ in \cref{eq:history_sequence}
from $k=0$ towards the core until, at $k=k_\star$, the condition
\begin{equation}
    Q_{\mathrm{EW}}^2\!\left(\Phi^{(k_\star)}\right)
    \ge
    \xi_{\mathrm{clust}}M_W^2
    \label{eq:cluster_condition}
\end{equation}
is first satisfied. Here $\xi_{\mathrm{clust}}\gtrsim 1$ is a
user-defined clustering-threshold parameter. The reconstructed Born-like object is
\begin{equation}
  \widehat\Phi_{\mathrm B}
  \equiv
  \Phi^{(k_\star)}.
  \label{eq:bornlike_object}
\end{equation}
This state is used to evaluate the EWSL, and the LHE weight is modified
according to \cref{eq:multiplicative_weight}. A larger
value of $\xi_{\mathrm{clust}}$ makes the stopping condition more
restrictive, so the replay proceeds further towards a lower-multiplicity
core; a smaller value stops the replay earlier. If no state in the history
 $\mathcal H$ satisfies \cref{eq:cluster_condition}, no EW Sudakov
correction is assigned and
the event retains its unmodified weight.

This design for the inclusion of the EWSL guarantees that the same
corrections are included just below and just above the FxFx merging
scale, and therefore the (in)dependence of the predictions on this
scale is unchanged.

This construction generalises the EWSL treatment of MC@NLO predictions
in \mgamc{}~\cite{Pagani:2023wgc}. In that
implementation, when an $\mathbb H$ event fails the condition in
\cref{eq:sudakov_domain}, its associated underlying-Born $\mathbb S$
configuration is used to evaluate the EWSL correction. This corresponds
effectively to one clustering step.

\subsection{Resonance-aware density-matrix reweighting}
\label{sec:resonance_aware_ewsl}
\label{sec:particles}
\label{sec:modes}
\label{sec:density_matrices}
\label{sec:ewsl_weight}
\label{sec:rest_independence}
\label{sec:multi_resonance}

If the clustering exposes a massive unstable resonance, such as a top
quark or a $W^\pm$ or $Z$ boson, the scalar method outlined so far
does not capture the helicity dependence of the EWSL correction.
Although the decay-product kinematics remain in the event, they do not
enter the factor $w\!\left(\widehat\Phi_{\mathrm B}\right)$. Since
the EWSL depend on the helicity state of the resonance, this can lead
to an incorrect reweighting of production-decay correlations. We
address this issue using resonance helicity density matrices, building
on the implementation of ref.~\cite{Durupt:2025wuk}.

For a general Born process with $N_{\mathrm R}$ on-shell resonances, the spin
density matrix can be defined as follows. Let
$a$ denote all production degrees of freedom other than the
resonance helicities, and let $b_r$ denote the degrees of freedom
summed over in the decay of resonance $r$. The Born production
matrix is
\begin{equation}
  \Bmat_{\lambda_1\cdots\lambda_{N_{\mathrm R}};\,
  \lambda_1'\cdots\lambda_{N_{\mathrm R}}'}
  =
  \sum_{a}
  \MB_{\lambda_1\cdots\lambda_{N_{\mathrm R}},a}\,
  \MB_{\lambda_1'\cdots\lambda_{N_{\mathrm R}}',a}^{*}\,.
  \label{eq:B_def}
\end{equation}
For each resonance, the decay matrix is defined with its indices ordered as
\begin{equation}
  \Cmat^{(r)}_{\lambda_r';\lambda_r}
  =
  \sum_{b_r}
  \mathcal D^{(r)}_{\lambda_r b_r}\,
  \mathcal D^{(r)*}_{\lambda_r'b_r}.
  \label{eq:C_single}
\end{equation}
Here $\mathcal D^{(r)}$ denotes the Born decay amplitude of resonance $r$.
The full decay matrix is
\begin{equation}
  \Cmat
  =
  \bigotimes_{r=1}^{N_{\mathrm R}}\Cmat^{(r)},
  \qquad
 \Cmat_{\lambda_1'\cdots\lambda_{N_{\mathrm R}}';\,
 \lambda_1\cdots\lambda_{N_{\mathrm R}}}
  =
  \prod_{r=1}^{N_{\mathrm R}}
  \Cmat^{(r)}_{\lambda_r';\lambda_r}.
  \label{eq:C_tensor}
\end{equation}
The spin-correlated Born contribution is proportional to
\begin{align}
  \tr[\Bmat\,\Cmat]
  &=
  \sum_{\lambda_1,\dots,\lambda_{N_{\mathrm R}}}
  \sum_{\lambda_1',\dots,\lambda_{N_{\mathrm R}}'}
  \Bmat_{\lambda_1\cdots\lambda_{N_{\mathrm R}};\,
  \lambda_1'\cdots\lambda_{N_{\mathrm R}}'}
  \,
  \Cmat_{\lambda_1'\cdots\lambda_{N_{\mathrm R}}';\,
  \lambda_1\cdots\lambda_{N_{\mathrm R}}}\nonumber\\
  &\equiv
  \sum_{\lambda}
  \sum_{\lambda'}
  \Bmat_{\lambda;\,\lambda'}
  \,
  \Cmat_{\lambda';\,\lambda},
  \label{eq:trace_lambda}
\end{align}
where we have introduced the shorthand notation $\lambda$ and
$\lambda'$ in the second line. The tensor-product form in
\cref{eq:C_tensor} assumes
non-overlapping decay systems, with no resonance included as both a
parent and a descendant in separate factors. A nested cascade, such as
$t\to bW^+\to b\ell^+\nu_\ell$, requires recursive contraction over the
intermediate-resonance spin indices. The final
contracted production-decay result is independent of the spin
basis provided that the production and decay matrices are transformed
consistently; the individual matrices are basis-dependent. Because the
EWSL correction is applied to the production part alone, we use the
helicity basis, which is natural in the high-energy regime specified
in \cref{eq:sudakov_domain_strict}.

A resonance requiring density treatment can enter the EWSL event
weight in two ways.
\begin{itemize}
\item \emph{Case 1: decays generated in the hard matrix element.} The
  decay products are ordinary external legs of a single hard
  amplitude. The assigned clustering history\footnote{For such
    processes, we recommend setting $\xi_{\mathrm{clust}}$, see
    \cref{eq:cluster_condition}, to a value larger than unity such
    that decay products of the (off-shell) Z or top quarks will be
    clustered before the EW Sudakov corrections are applied.}
  determines whether a candidate subsystem is promoted to a resonance:
  \begin{itemize}
  \item If the history leaves the subsystem resolved, its decay
    products remain external legs of $\widehat\Phi_{\mathrm B}$ and
    the scalar Sudakov path is used.
  \item If the history assigns the subsystem to a pole associated with
    an unstable massive particle, the subsystem is projected to an
    on-shell resonance and the density-matrix reweighting is used.
  \item If the assigned mother is not density-treated, for example a
    photon or a QCD parton, no resonance density matrix is introduced
    and the scalar path is retained.
  \end{itemize}
  For a hard amplitude such as $pp\to\ell^+\ell^-+X$, which contains
  both $\gamma^*$ and $Z$ contributions, the clustering determines\footnote{In
    the FxFx clustering, the parent assignment is determined by the
    integration channel to which the event belongs.}
  whether the subsystem is assigned to a resonant $Z$ configuration,
  for which the density-matrix reweighting is used, or to a non-resonant
  $\gamma^*$ configuration, for which the scalar path is retained.

\item \emph{Case 2: on-shell production followed by MadSpin decay.}
  The hard process contains an explicit on-shell resonance and
  MadSpin~\cite{Artoisenet:2012st} attaches a decay in the
  narrow-width approximation. The mother-daughter assignment is then
  part of the event record. The recorded decay chain is reversed
  before the remaining FxFx QCD history is traversed. The resulting
  resonance therefore always belongs to the Born-like production
  object, and the density-matrix reweighting is used, with its MadSpin decay products
  entering the decay density matrix.
\end{itemize}

For events requiring density-matrix reweighting, we evaluate the EWSL
correction to the production amplitude at the clustered Born-like
configuration $\widehat\Phi_{\mathrm B}$. As in
\cref{eq:B_def,eq:trace_lambda}, $\lambda$ collectively denotes the
resonance helicities and $a$ the remaining production degrees of
freedom. For fixed $\lambda$ and $a$, we write
\begin{equation}
\MB_{\lambda,a}\longrightarrow
\MB_{\lambda,a}+\delta\MB_{\lambda,a}
\equiv
\left(1+\Delta_{\lambda,a}\right)\MB_{\lambda,a},
\quad
\Delta_{\lambda,a}=\mathcal O(\alpha)
\ \ \text{(generally complex)},
\label{eq:amp_level_Delta}
\end{equation}
where the gauge-group indices of \cref{eq:la_factorised} are
suppressed. The EWSL correction preserves the external helicities,
while its magnitude and phase can depend on both $\lambda$ and $a$.
Substituting \cref{eq:amp_level_Delta} into the production matrix of
\cref{eq:B_def} and retaining terms through $\mathcal O(\alpha)$ gives
\begin{align}
&\sum_{a}\left(\MB_{\lambda,a}+\delta\MB_{\lambda,a}\right)
\left(\MB_{\lambda',a}+\delta\MB_{\lambda',a}\right)^{*}
\nonumber\\
&\quad=
\Bmat_{\lambda;\lambda'}
+\sum_a
\left(\Delta_{\lambda,a}+{\Delta_{\lambda',a}}^{*}\right)
\MB_{\lambda,a}\MB_{\lambda',a}^{*}
+\mathcal O(\alpha^2).
\label{eq:B_ewsl_rest_dependent}
\end{align}

We use a factorised prescription to simplify the implementation and
reduce the amount of correction data stored and passed between the
Sudakov and density-matrix modules. The Sudakov module sums over $a$
at fixed $\lambda$ and supplies one complex correction factor per
resonance-helicity configuration, while the density-matrix module
provides the full Born matrix. Combining these quantities avoids
evaluating separate Sudakov interference terms for each pair of
distinct resonance helicities. We define the correction factor by the
Born-weighted average over $a$,
\begin{equation}
\Delta_\lambda
=
\frac{
  \sum_a \Delta_{\lambda,a}\lvert\MB_{\lambda,a}\rvert^2
}{
  \sum_a \lvert\MB_{\lambda,a}\rvert^2
}.
\label{eq:Delta_born_average}
\end{equation}
The resulting factorised update of the production matrix is
\begin{equation}
\Bmat^{\mathrm{EWSL}}_{\lambda;\lambda'}
=
\Bmat_{\lambda;\lambda'}
+\left(\Delta_{\lambda}+{\Delta_{\lambda'}}^{*}\right)
\Bmat_{\lambda;\lambda'}
+\mathcal O(\alpha^2).
\label{eq:B_ewsl_linearised}
\end{equation}

By construction, this prescription reproduces the diagonal entries of
\cref{eq:B_ewsl_rest_dependent} exactly at $\mathcal O(\alpha)$. The
off-diagonal entries describe interference between different resonance
helicities and involve the products
$\MB_{\lambda,a}\MB_{\lambda',a}^{*}$. These interference weights
differ from the squared amplitudes used in
\cref{eq:Delta_born_average}. Thus, for a fixed off-diagonal element
$(\lambda,\lambda')$, the factorised update is generally approximate
when either $\Delta_{\lambda,a}$ or $\Delta_{\lambda',a}$ varies with
the helicities of the other external particles summed over in $a$.

Finally, the corrected production matrix is contracted with the decay
matrix. The uncorrected
spin-correlated scalar is $\tr[\Bmat\,\Cmat]$, while its reweighted
counterpart is $\tr[\Bmat^{\mathrm{EWSL}}\Cmat]$. Let
$\Delta_{\lambda;\lambda'}=\Delta_\lambda
\delta_{\lambda\lambda'}$ define the corresponding diagonal matrix.
The ratio of the two contractions gives the event-reweighting factor,
\begin{equation}
w =
\frac{\tr\!\left[\Bmat^{\mathrm{EWSL}}\,\Cmat\right]}
     {\tr\!\left[\Bmat\,\Cmat\right]}
=
1
+
2\,\Re\!\left(
\frac{\tr\!\left[\Delta\,\Bmat\,\Cmat\right]}{\tr\!\left[\Bmat\,\Cmat\right]}
\right)
+\mathcal O(\alpha^2).
\label{eq:w_lambda}
\end{equation}

\section{Validation and numerical results}
\label{sec:validation}

We validate the reweighting in two steps. The first test uses
$Z$-boson production with matrix-element jets,
$pp\to Z+\mathrm{jets}$, and probes the FxFx clustering
path introduced in \cref{sec:baseline}. The $Z$ boson is generated
on shell and kept stable, so no decay reconstruction, lepton dressing
or issues associated with lepton-level QED radiation enter the
comparison. This provides a clean test of the reconstruction of the
Born-like object and of the multiplicative weak-Sudakov weight.
The second test, discussed in
\cref{sec:ww_validation}, uses fully leptonic $W^+W^-$ production
and validates the resonance-aware density-matrix reweighting.

The implementations of density-matrix reweighting Cases~1 and~2 use essentially the
same code, except that in Case~2 the first clustering steps are forced
to reconstruct the resonances recorded in the event. We present
resonance-validation results only for Case~2, since on-shell
production followed by MadSpin decays allows a more straightforward
comparison with exact fixed-order NLO electroweak corrections to the
production process. 

Using \mgamc{}~\cite{Alwall:2014hca,Frederix:2018nkq}, we generate all
event samples and fixed-order references at $\sqrt{s}=13\TeV$. The fixed-order NLO
calculations use the standard \textsc{MadFKS}
subtraction~\cite{Frixione:1995ms,Frederix:2009yq} and
\textsc{MadLoop} one-loop infrastructure~\cite{Hirschi:2011pa}, while
the EWSL reweighting uses the \mgamc{} electroweak Sudakov
module~\cite{Pagani:2021vyk,Pagani:2023wgc} and the spin-density-matrix
module~\cite{Durupt:2025wuk}. The events are showered with
\textsc{Pythia}~8~\cite{Bierlich:2022pfr}.

\subsection{\texorpdfstring{$Z+\mathrm{jets}$}{Z+jets}: FxFx clustering validation}
\label{sec:zjets_validation}

The $Z+\mathrm{jets}$ sample tests the part of the algorithm that reconstructs
the Born-like object from an FxFx clustering history before evaluating
the Sudakov logarithms. Since the $Z$ is stable, it remains an
ordinary external leg of the reconstructed hard process and no
resonance density matrix is introduced. The relevant comparison is
therefore between a showered FxFx prediction with and without the
EWSL weight.

For an observable $x$, denote the FxFx prediction by
$\NLOQCD+\mathrm{PS}$. In the ratio
\begin{equation}
  \frac{
    \dd\sigma\!\left[
      (\NLOQCD+\mathrm{PS})\otimes\EWSL
    \right]/\dd x
  }{
    \dd\sigma\!\left[
      \NLOQCD+\mathrm{PS}
    \right]/\dd x
  }
  \equiv \KEWSL,
  \label{eq:nlops_correction}
\end{equation}
the numerator and denominator use the same NLO QCD hard process and
QCD shower settings, reducing the sensitivity of the ratio to QCD
effects. In the high-energy region, the ratio is therefore expected to
track the logarithmically enhanced part of the fixed-order electroweak
$K$-factor. We compare it with both the additive fixed-order (FO) ratio,
\begin{equation}
  \left.
  \frac{
    \dd\sigma\!\left[
      \NLOQCD+\NLOEW
    \right]/\dd x
  }{
    \dd\sigma\!\left[
      \NLOQCD
    \right]/\dd x
  }
  \right|_{\mathrm{FO}}
  \equiv \KFO,
  \label{eq:nlo_correction}
\end{equation}
and the LO-normalised electroweak ratio used in the multiplicative
combination,
\begin{equation}
  \left.
  \frac{
    \dd\sigma\!\left[
      \LOQCD+\NLOEW
    \right]/\dd x
  }{
    \dd\sigma\!\left[
      \LOQCD
    \right]/\dd x
  }
  \right|_{\mathrm{FO}}
  \equiv \KFOLO,
  \label{eq:lo_correction}
\end{equation}
where $\NLOQCD$ denotes the LO QCD result plus the NLO QCD correction.
In this compact additive notation, $\NLOEW$ denotes the electroweak
correction alone rather than a complete prediction.
Because the EWSL factor multiplies the full hard-event weight, the
implementation realises a multiplicative combination of QCD and EW
corrections, cf.~\cref{eq:multiplicative_weight}. The corresponding
fixed-order prescription multiplies the NLO QCD prediction by
$\KFOLO$, which therefore provides the natural benchmark for
$\KEWSL$ in the Sudakov regime.

We focus on two observables that probe complementary aspects of the
FxFx clustering prescription. The $Z$-boson transverse momentum,
$\pT(Z)$, directly probes the hard electroweak recoil. By contrast, a
large leading-jet transverse momentum, $\pT(j_1)$, can arise from a
hard dijet configuration containing a soft or collinear $Z$, for which
the weak correction remains small. Jets are reconstructed with the
anti-$k_T$ algorithm, $R=0.4$ and $\pT>30\GeV$, clustering both QCD
partons and photons. A veto on a second jet suppresses dijet recoil
configurations and selects topologies in which the jet transverse
momentum also sets the scale of the weak Sudakov logarithms.

The FxFx sample is generated with up to two matrix-element jets,
\begin{center}
\code{p p > Z [QCD] @0; p p > Z j [QCD] @1; p p > Z j j [QCD] @2},
\end{center}
in the five-flavour scheme. We use the
\texttt{NNPDF31\_nlo\_as\_0118\_luxqed} parton distribution function (PDF)
set~\cite{Manohar:2016nzj,Bertone:2017bme} and the $G_\mu$ input
scheme. The FxFx run-card settings are \code{ickkw=3}, a
matrix-element jet cut $\pT^{\mathrm{ME}}=10\GeV$, $k_T$
generation-level jets with $R=1$, and the default dynamical
renormalisation and factorisation scales based on the FxFx clustering
history~\cite{Frederix:2012ps}. A single event sample is generated\footnote{We
use an $H_T^2$ sampling bias to improve statistics in the multi-TeV tails;
this bias is divided out of the stored event weights.} and
the FxFx merging veto is then applied at $Q_{\mathrm{cut}}=20$, $30$
and $40\GeV$ on a per-event basis; unless stated otherwise we quote the
$Q_{\mathrm{cut}}=30\GeV$ result. Because the three merging scales are
evaluated on the same events, their differences are statistically
correlated. We use
$\xi_{\mathrm{clust}}=1.5$ for the EWSL reweighting. The masses and
couplings are set to
\begin{align*}
  M_Z&=91.188\GeV,\quad
  M_W=80.419\GeV,\quad
  M_H=125\GeV,\\
  m_t&=173.3\GeV,\quad
  G_F=1.16639\times10^{-5}\GeV^{-2},
\end{align*}
with a diagonal Cabibbo--Kobayashi--Maskawa (CKM) matrix. All widths
are set to zero. The events are showered with \textsc{Pythia}~8, with
hadronisation and multiparton interactions switched off.

We keep the $Z$ boson stable using \code{23:mayDecay=off}. For the
unreweighted baseline, the QED shower is off; for the reweighted
sample it is switched on together with the $\SDKweak$ Sudakov weight,
as described in \cref{sec:sdkweak}. Because the $Z$ is stable and
photons are clustered into jets, QED-shower effects are expected to be
small for the observables considered here. Nevertheless, the ratio in
\cref{eq:nlops_correction} will highlight the combined effect of the
EWSL and QED shower, which is expected to be a suitable proxy for the
exact EW corrections.

Since we focus on the $\pT(Z)$ and $\pT(j_1)$ observables (with or
without a jet veto), the fixed-order predictions are obtained with the
command
\begin{center}
\code{generate p p > z j [QCD QED]}.
\end{center}
This run uses the same input parameters as the FxFx event samples,
except that it uses $H_T/2$ for the renormalisation and factorisation
scales. For this run we impose a generation cut
$\pT(Z)>10\GeV$, while the analysis-level jet definition is applied
only during histogramming.

We have also checked observables that explicitly require a second jet,
including $\pT(j_2)$, $M(j_1j_2)$, and $M(Zj_1j_2)$. They can
be obtained from the same FxFx sample, but their fixed-order comparison
requires independent $Z+jj$ calculations. Compared with the $Z+j$-type
spectra discussed above, these observables have an additional
complication: depending on the coupling orders included, the fixed-order
prediction can contain resonant weak-boson decays into the two jets.
Such contributions are not generated by EWSL reweighting of the leading
QCD sample. Moreover, these observables populate the far tails less
efficiently; even with the available FxFx statistics, their
uncertainties are too large to support conclusions beyond those
already drawn from $\pT(Z)$ and $\pT(j_1)$. We therefore do not
show them.

\begin{figure}[htbp]
  \centering
  \includegraphics[width=0.49\textwidth]{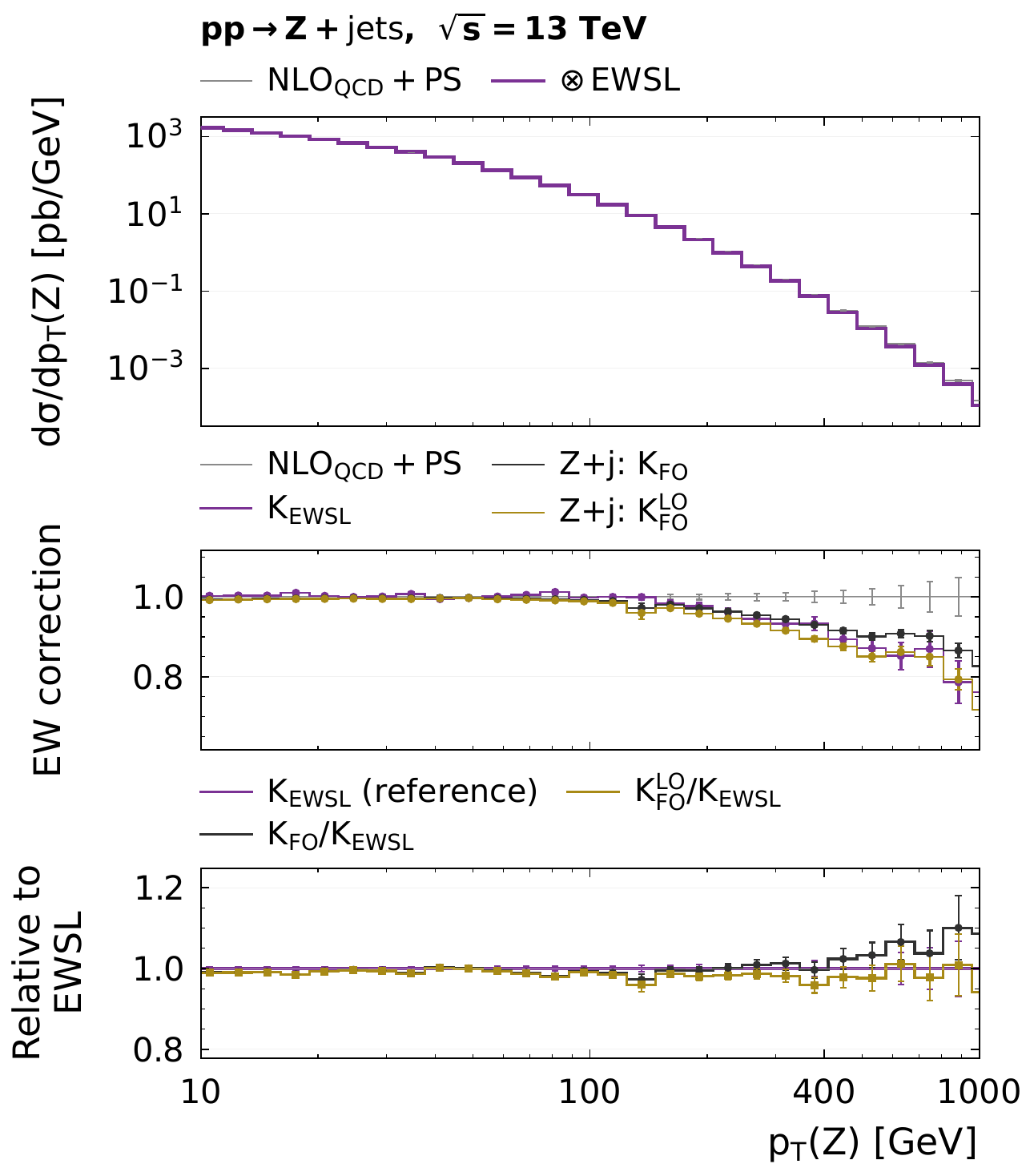}\hfill
  \includegraphics[width=0.49\textwidth]{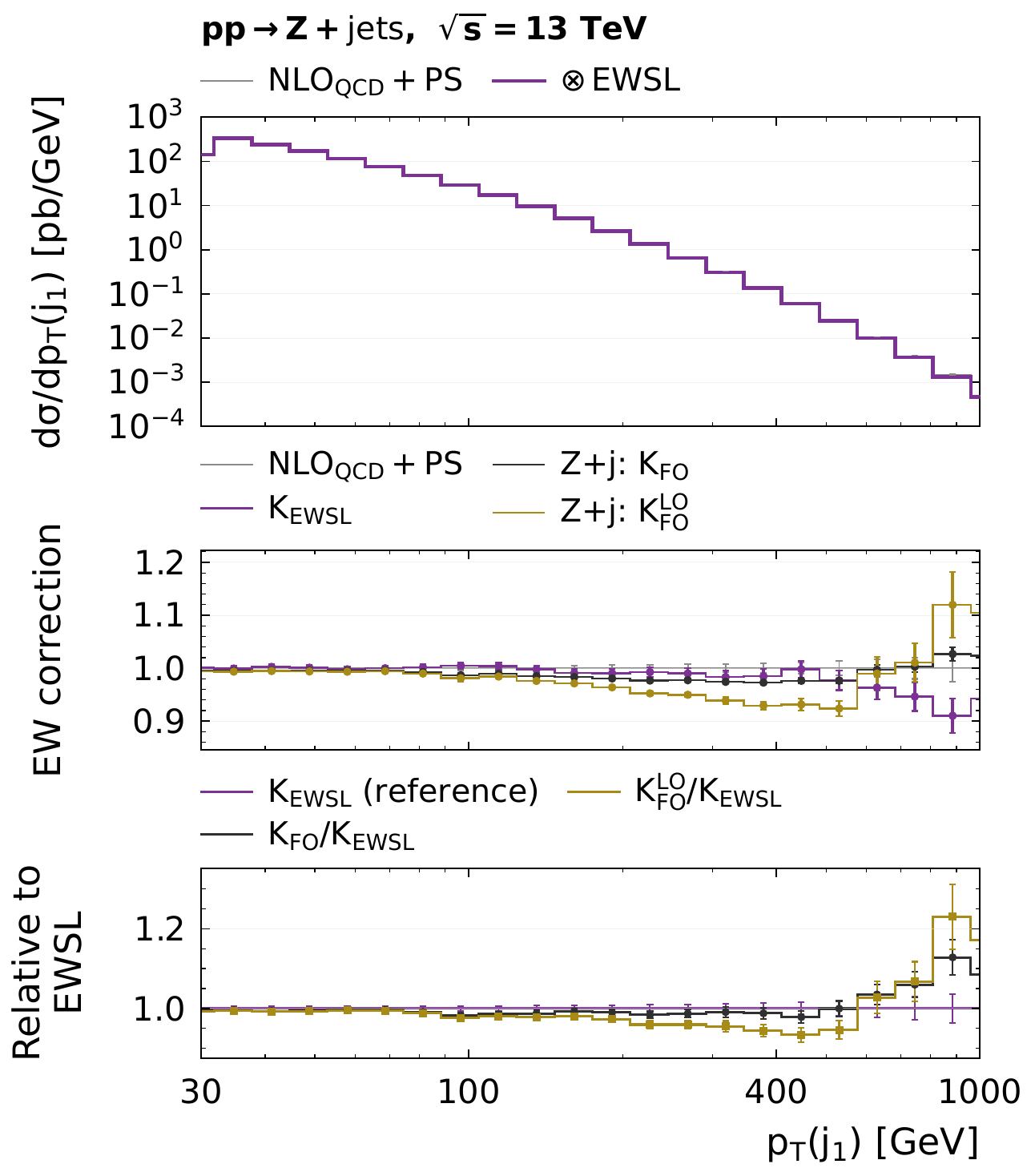}
  \caption{Validation of the EWSL reweighting on a $Z+\mathrm{jets}$ FxFx
    sample ($Q_{\mathrm{cut}}=30\GeV$, $\xi_{\mathrm{clust}}=1.5$):
    $\pT(Z)$ (left) and $\pT(j_1)$.
    The middle panels show $\KEWSL$ (purple), $\KFO$ (black) and $\KFOLO$
    (mustard), with the unreweighted grey reference at unity. The panels
    labelled ``Relative to EWSL'' show $\KFO/\KEWSL$ and
    $\KFOLO/\KEWSL$, with the purple $\KEWSL$ reference at unity;
    agreement means equal relative corrections.
    In the high-$\pT(Z)$ tail, the showered correction follows
    the multiplicative reference $\KFOLO$ of \cref{eq:lo_correction}
    more closely than the additive $\KFO$ of
    \cref{eq:nlo_correction}. Error bars are statistical.}
  \label{fig:ewsl_validation}
\end{figure}

\begin{figure}[htbp]
  \centering
  \includegraphics[width=0.49\textwidth]{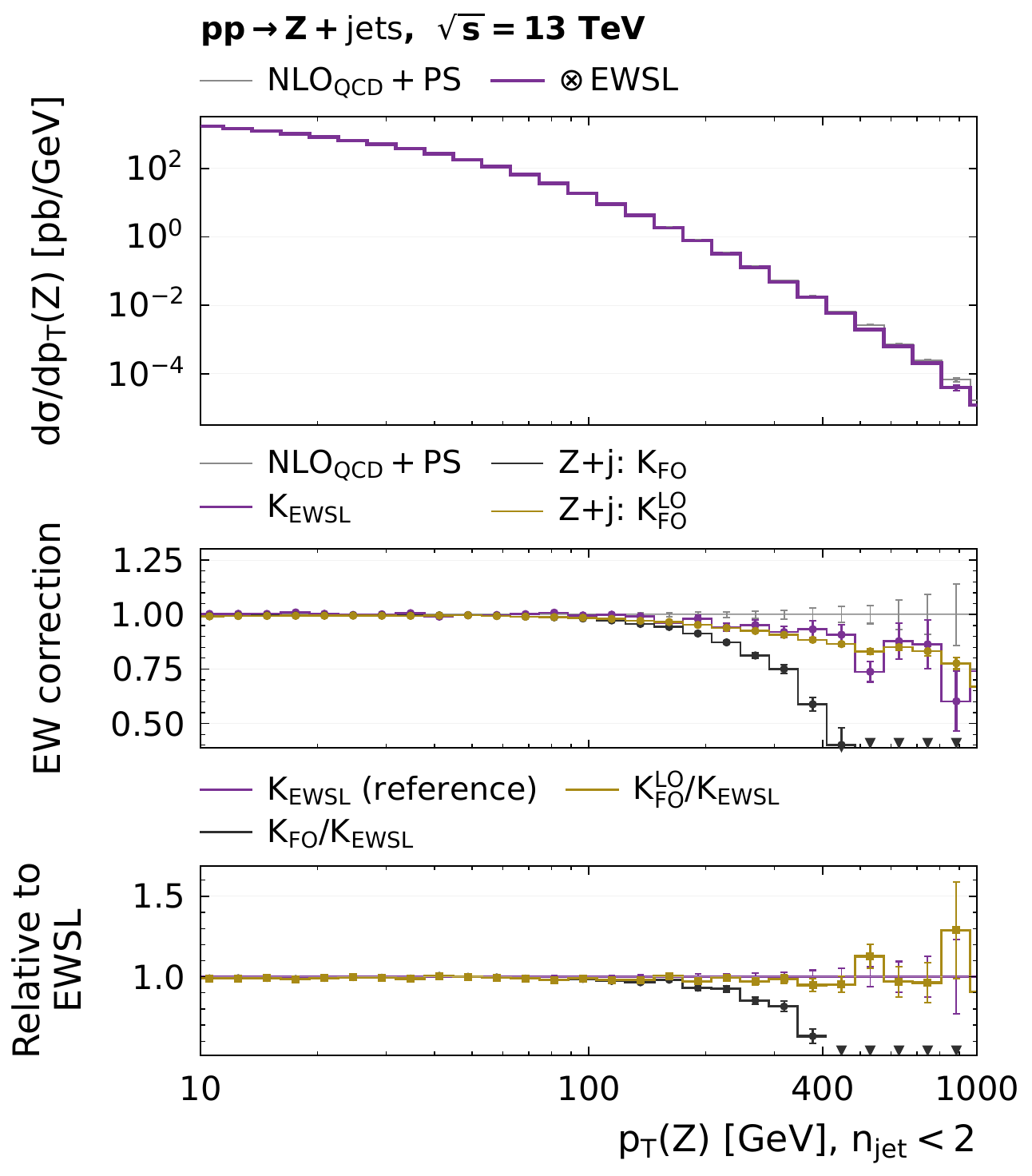}\hfill
  \includegraphics[width=0.49\textwidth]{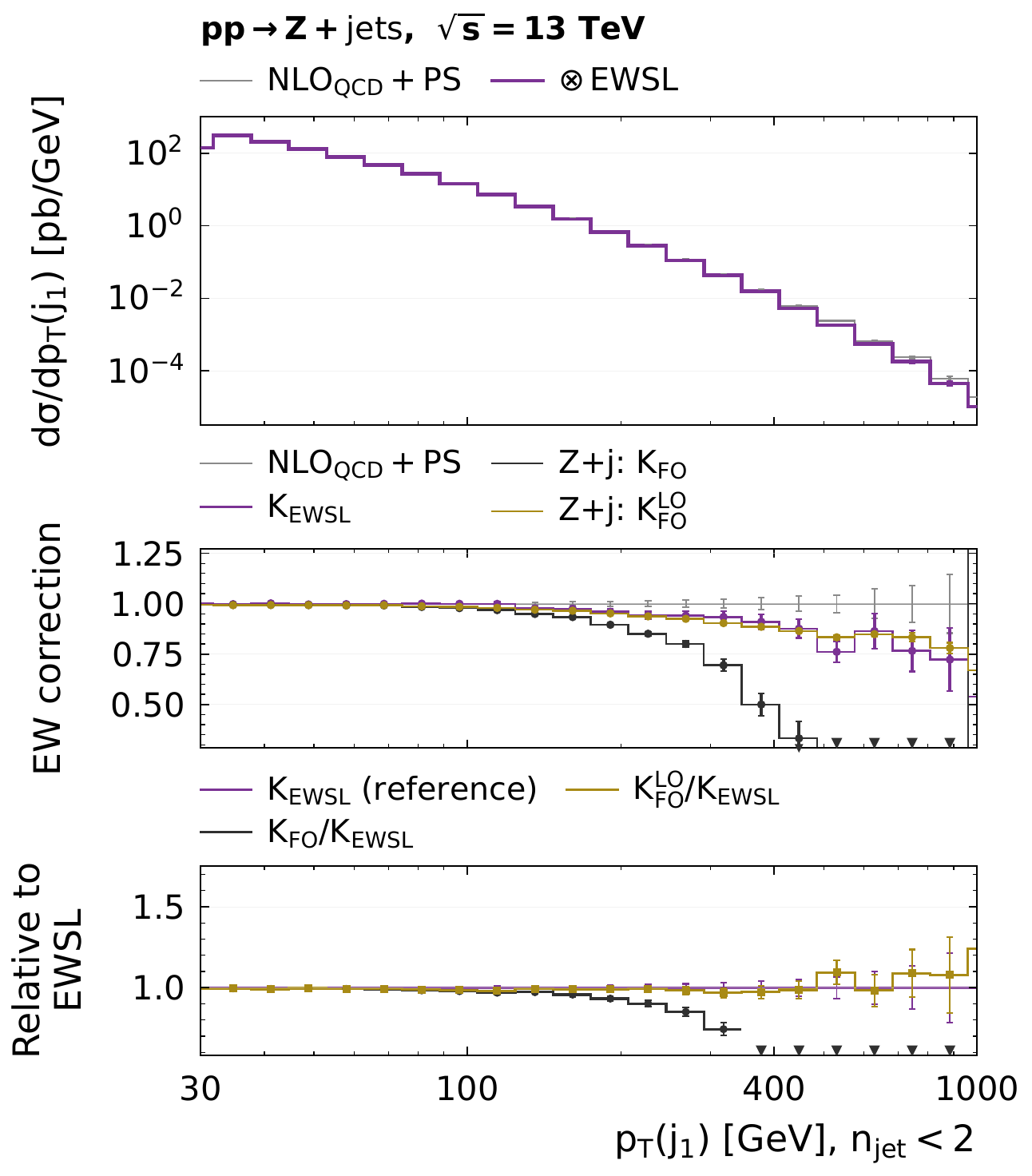}
  \caption{Jet-vetoed observables for the same $Z+\mathrm{jets}$ sample
    as in \cref{fig:ewsl_validation}:
    $\pT(Z)$ (left) and $\pT(j_1)$ (right) with
    $n_{\mathrm{jet}}<2$. The middle panels show $\KEWSL$, $\KFO$ and
    $\KFOLO$; the ``Relative to EWSL'' panels show $\KFO/\KEWSL$
    and $\KFOLO/\KEWSL$,
    with colours and line styles as in \cref{fig:ewsl_validation}.
    The veto suppresses hard QCD real-emission
    configurations, so the relevant fixed-order benchmark is $\KFOLO$,
    which the $\otimes\EWSL$ prediction closely tracks. $\KFO$ is
    destabilised by the veto, which drives its fixed-order denominator
    towards zero. Triangles just inside the frame mark off-scale central
    values; arrowheads on the frame mark truncated error bars.
    Error bars are statistical.}
  \label{fig:ewsl_jetveto}
\end{figure}

In \cref{fig:ewsl_validation,fig:ewsl_jetveto}, each main panel contains
two curves: the unreweighted FxFx prediction at $\NLOQCD+\mathrm{PS}$
in grey and the EWSL-reweighted FxFx prediction in purple. The
fixed-order predictions enter only through the ratio panels. The first
ratio panel shows the relative electroweak corrections $\KEWSL$, $\KFO$ and
$\KFOLO$ defined in
\cref{eq:nlops_correction,eq:nlo_correction,eq:lo_correction}, in
purple, black and mustard, respectively. The grey unity reference is
labelled $\NLOQCD+\mathrm{PS}$. The label $Z+j$ identifies the
fixed-order reference process.
In both figures, the panel labelled ``Relative to EWSL'' shows
$\KFO/\KEWSL$ and $\KFOLO/\KEWSL$, separating the size of the
electroweak corrections from their mutual agreement. Its purple unity
line is labelled $\KEWSL$ (reference), identifying the showered
EWSL correction used as the denominator.
Error bars show statistical uncertainties.

We begin with the inclusive $\pT(Z)$ spectrum shown in the left panel
of \cref{fig:ewsl_validation}. The showered EWSL ratio follows the
high-energy suppression seen in the fixed-order electroweak ratios.
The correction is close to unity below $\pT(Z)\simeq100\GeV$, where the hard invariants are not yet
parametrically above the weak scale, and then develops the expected
logarithmic suppression. For $\pT(Z)$ the effect reaches
approximately $-5\%$ at $300\GeV$ and about $-15\%$ at $1\TeV$.

The three correction factors remain close at moderate $\pT(Z)$.
In the high-$\pT(Z)$ tail, $\KEWSL$ follows the multiplicative
reference $\KFOLO$ more closely, while the additive ratio $\KFO$
gives a smaller suppression. This behaviour is consistent with the
multiplicative application of the EWSL correction to the QCD event
weights.

The right plot of \cref{fig:ewsl_validation}, which shows the
$\pT(j_1)$ distribution, exhibits much smaller EW corrections. For
both fixed-order ratios and for the EWSL-reweighted FxFx result, the
magnitude of the EW correction remains below $10\%$ over the full
range considered, without the usual increase in logarithmic
suppression towards large $\pT(j_1)$. At large leading-jet transverse
momenta, a significant contribution comes from configurations with a
second hard jet and a relatively soft or collinear $Z$ boson. The
presence of the latter reduces the size of the EW correction.

To further test this interpretation, \cref{fig:ewsl_jetveto} shows the
same distributions with a veto on the second jet. Both $\pT(Z)$ and
$\pT(j_1)$ now exhibit the characteristic EW suppression dominated
by the EWSL\@. The correction factor in the FxFx-merged result is almost
identical to the LO-normalised fixed-order EW correction $\KFOLO$. This
agreement supports the conclusion that, once the veto suppresses hard dijet
configurations with a soft $Z$ boson, the clustering replay
identifies the hard electroweak topology relevant to the Sudakov
correction.

\begin{figure}[htbp]
  \centering
  \includegraphics[width=0.99\textwidth]{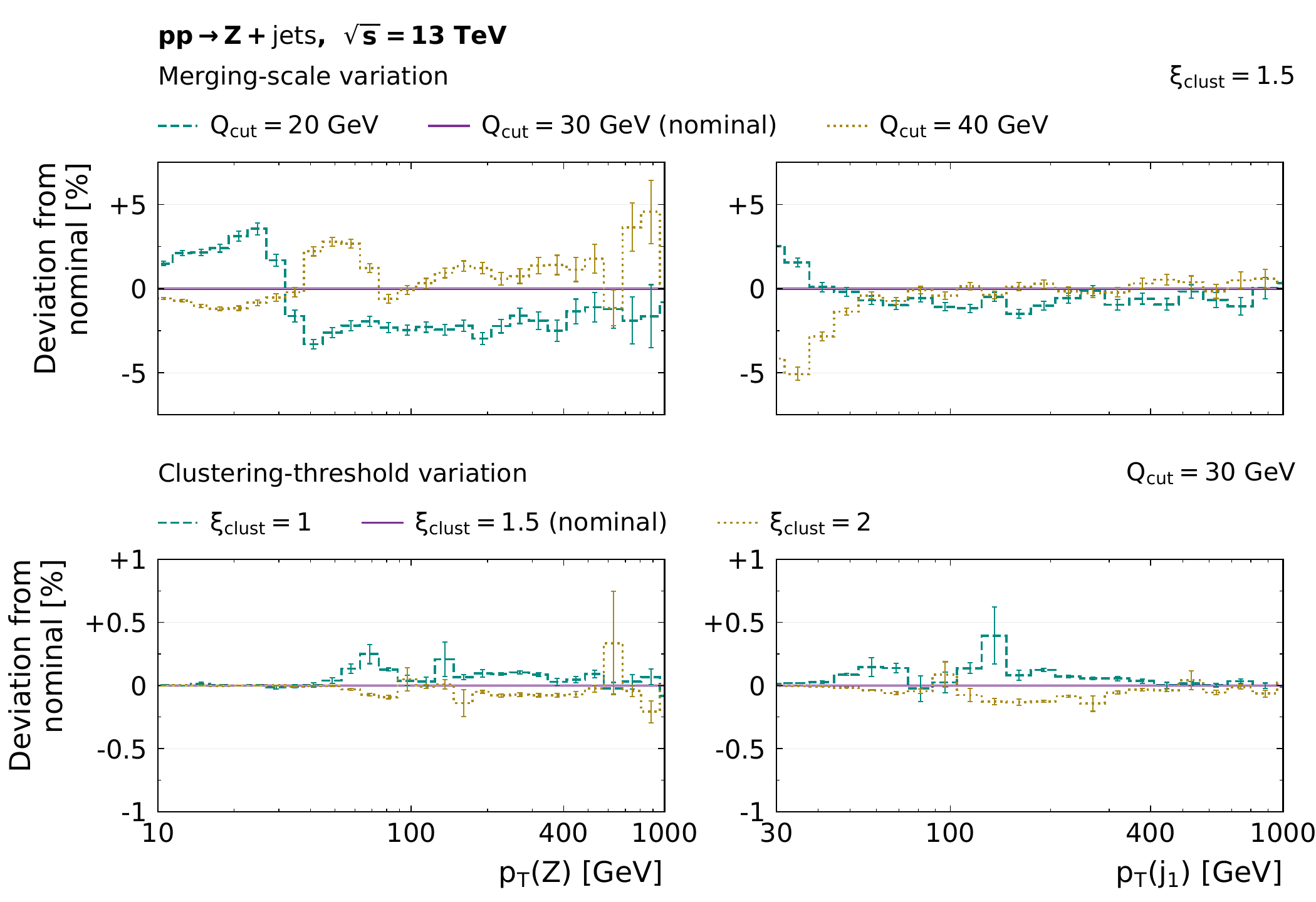}
  \caption{Parameter dependence of the EWSL-reweighted predictions for
    $\pT(Z)$ (left) and $\pT(j_1)$ (right), shown as percentage
    deviations from the nominal reweighted prediction at
    $Q_{\mathrm{cut}}=30\GeV$, $\xi_{\mathrm{clust}}=1.5$.
    Top row: variation of the FxFx merging scale
    $Q_{\mathrm{cut}}\in\{20,30,40\}\GeV$ at
    $\xi_{\mathrm{clust}}=1.5$, including the ordinary QCD merging
    dependence. Bottom row: variation of the clustering threshold
    $\xi_{\mathrm{clust}}\in\{1,1.5,2\}$ at
    $Q_{\mathrm{cut}}=30\GeV$. The purple zero line is the nominal
    prediction. Each row shares a legend and a common vertical scale;
    the two rows use different scales. Statistical error bars include
    the measured same-event covariance with the nominal prediction.}
  \label{fig:ewsl_robustness}
\end{figure}

Finally, \cref{fig:ewsl_robustness} tests the dependence on the two
auxiliary parameters entering this validation for the same two
observables: the FxFx merging scale $Q_{\mathrm{cut}}$ and the
clustering-threshold parameter $\xi_{\mathrm{clust}}$ of
\cref{eq:cluster_condition}.  The $Q_{\mathrm{cut}}$ variation
probes the ordinary merging uncertainty of the underlying QCD sample,
while $\xi_{\mathrm{clust}}$ changes only the electroweak
reconstruction step.

All four panels show
\begin{equation*}
  \left[
    \frac{\dd\sigma^{\mathrm{EWSL}}_{\mathrm{var}}/\dd x}
         {\dd\sigma^{\mathrm{EWSL}}_{\mathrm{nom}}/\dd x}
    -1\right],
\end{equation*}
the deviation from the nominal EWSL-reweighted prediction
at $Q_{\mathrm{cut}}=30\GeV$ and $\xi_{\mathrm{clust}}=1.5$.
The errors retain the measured covariance between each variation and
the nominal prediction; no independent error is assigned to the
nominal prediction divided by itself.

In the upper row, the teal dashed, purple solid and mustard dotted
curves correspond to $Q_{\mathrm{cut}}=20$, $30$ and $40\GeV$,
respectively, at fixed $\xi_{\mathrm{clust}}=1.5$. This comparison
retains the ordinary merging dependence of the underlying QCD sample.
The few-per-cent shifts near the merging scale and in the tails are
consistent with the usual FxFx systematics. 

The lower row isolates the dependence on $\xi_{\mathrm{clust}}$ at
fixed $Q_{\mathrm{cut}}=30\GeV$. Teal, purple and mustard correspond
to $\xi_{\mathrm{clust}}=1$, $1.5$ and $2$, respectively.  These are
pure reweighting-side variations of the same events.  Their shifts
remain sub-per-cent over the tested range, as resolved by the narrower
common vertical scale of this row.

\subsection{Density-matrix reweighting validation: \texorpdfstring{$W^+W^-$}{WW} with leptonic decays}
\label{sec:ww_validation}

We consider fully leptonic $W^+W^-$ production,
\begin{equation*}
  pp\to W^+W^-\to\mu^+\nu_\mu\,e^-\bar\nu_e,
\end{equation*}
at $\sqrt{s}=13\TeV$. The showered samples are generated at NLO
QCD\footnote{The EWSL machinery uses the clustering history $\mathcal
H$ provided by the FxFx implementation, so FxFx needs to be enabled
when generating this sample. However, only one hard-process
multiplicity is included: no samples of different multiplicities are
merged, and neither the FxFx $\alpha_s$ reweighting nor its QCD
Sudakov suppression is applied, and therefore, without EWSL
reweighting, the events are statistically equivalent to a default
NLO+PS sample generated with FxFx disabled.}; the nominal
clustering-threshold parameter of \cref{eq:cluster_condition} is
$\xi_{\mathrm{clust}}=1$. The two $W$ bosons are produced on shell and
decayed with MadSpin~\cite{Artoisenet:2012st} in the narrow-width
approximation. Unless stated otherwise, the MadSpin decay retains the
full production-decay spin correlations. The different-flavour final
state makes the assignment of each charged lepton and neutrino to its
parent $W$ unambiguous. This is Case~2 of \cref{sec:modes}: the two
spin-one resonances span a nine-dimensional joint helicity space, so
the production density matrix is a $9\times9$ matrix.

The hard process is generated in the four-flavour scheme (4FS), with
massive bottom quarks, no bottom-quark parton density, and real
bottom-quark emission omitted. This defines a top-quark-free $W^+W^-$
signal.\footnote{For
consistency with the $Z+\text{jets}$ study
we use the same PDF set, whose fit employs five active flavours, while
excluding bottom quarks from the hard-process parton definition. This
formally mixed choice is not expected to affect the correlated validation
tests materially.}

The treatment of photon-initiated channels requires a separate
clarification. The
photon-induced Born channel $\gamma\gamma\to W^+W^-$ is of order
$\alpha^2$, just like the quark-induced Born process
$q\bar q\to W^+W^-$. We nevertheless exclude it from the central
validation, because the EWSL-reweighted sample contains only
quark-induced hard events and does not generate a new
photon-initiated channel. These photon-induced contributions are
currently not supported in an $\NLOQCD+\mathrm{PS}$ simulation.
Including such events would test a different
partonic mixture and would partly compensate the negative weak
Sudakov correction. In the fixed-order calculation the photon is
retained as an initial-state parton so that all QED subtraction terms are
generated. We then show results both with the physical photon PDF and
with its density set to zero; the latter is the matched
reference for the EWSL samples. The full photon-PDF result is retained
in the plots to display the size of the omitted photon-induced
contribution, rather than treating it as a higher-order effect.

We compare three showered predictions derived from the same NLO QCD
hard process:
\begin{itemize}
  \item the \emph{baseline}, with spin-correlated MadSpin decays and
    no EWSL reweighting;
  \item the \emph{density} result, in which the EWSL correction is
    applied to the production density matrix and contracted with the
    MadSpin decay matrices according to \cref{eq:w_lambda}; and
  \item the \emph{scalar} result, in which the polarisation-summed
    EWSL factor is applied to the on-shell production process before
    the MadSpin decay.
\end{itemize}

The scalar reweighting preserves the spin correlations already present
in the baseline event, but it cannot change them according to the
helicity dependence of the electroweak correction. This corresponds to
the treatment of production and decay used in
ref.~\cite{Pagani:2023wgc}. The density result does include this
dependence. Thus their difference isolates precisely the effect that
the density-matrix construction is designed to capture. The density
and scalar samples use a QCD$+$QED shower, whereas the baseline uses
a QCD-only shower. Consequently, the density-scalar comparison uses
identical shower settings, while ratios to the baseline may also
contain QED-shower effects at lepton level. The density matrix and
EWSL weight themselves are evaluated on the common pre-shower LHE
kinematics. The charged-lepton distributions use bare, post-shower
leptons, without photon recombination. The $W$ distributions use the
Monte Carlo truth momenta, including any photon radiation in the
decays or from the $W$ bosons themselves. The fixed-order reference
uses on-shell, undecayed $W$ bosons.

For observables constructed from the $W$ momenta, the narrow-width
approximation therefore permits a direct comparison of the NLO+PS
predictions with fixed-order on-shell $W^+W^-$ production. We use
$H_T/2$ as the fixed-order renormalisation and factorisation scale and
compare with the two correction factors defined in
\cref{eq:nlo_correction,eq:lo_correction}. We denote the showered
correction factors relative to the unreweighted baseline by
$K_{\mathrm{scalar}}$ and $K_{\mathrm{density}}$, following
\cref{eq:nlops_correction} for each prescription.

\begin{figure}[htbp]
  \centering
  \includegraphics[width=0.475\textwidth]{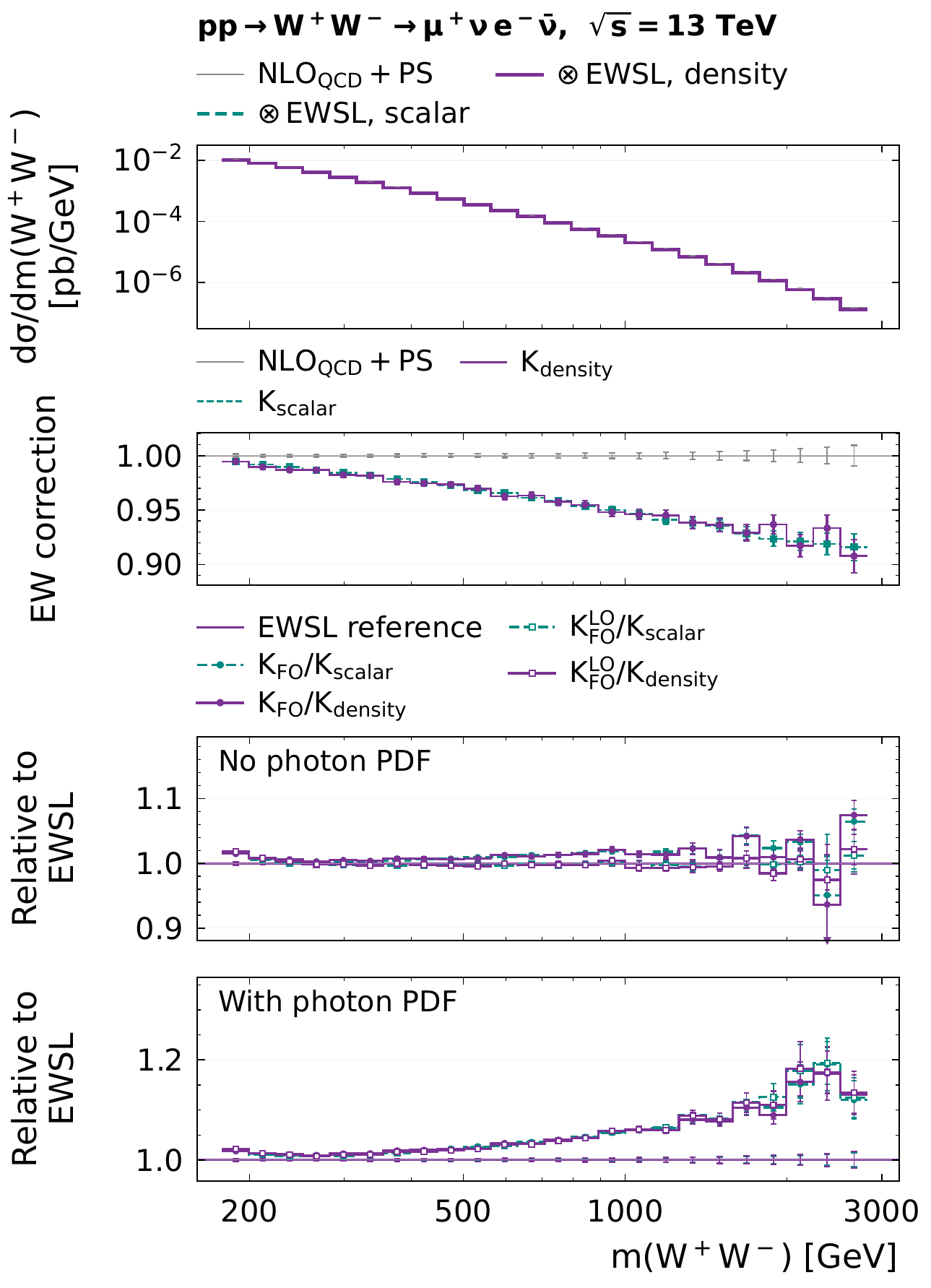}\hfill
  \includegraphics[width=0.475\textwidth]{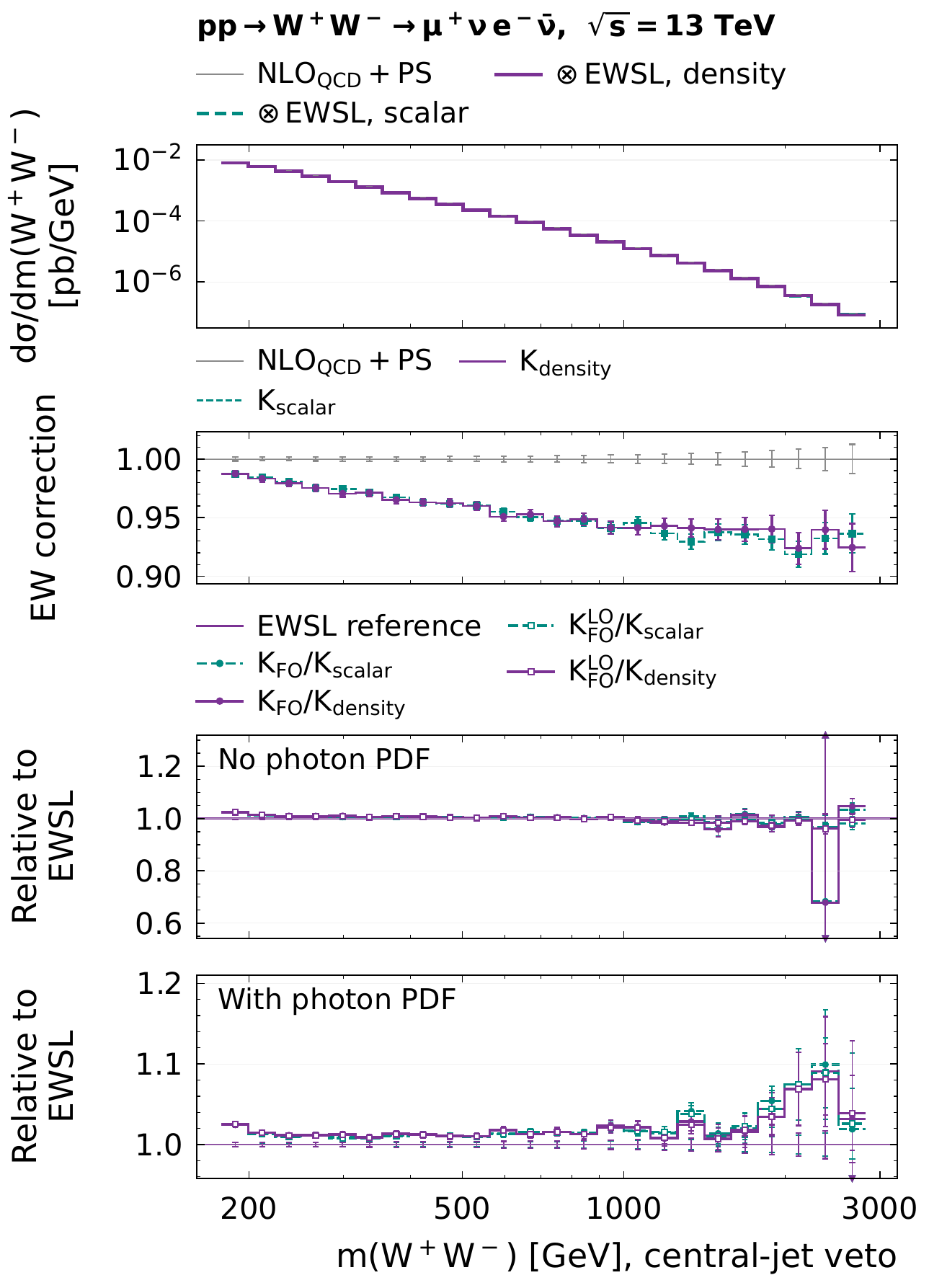}\\[1pt]
  \includegraphics[width=0.475\textwidth]{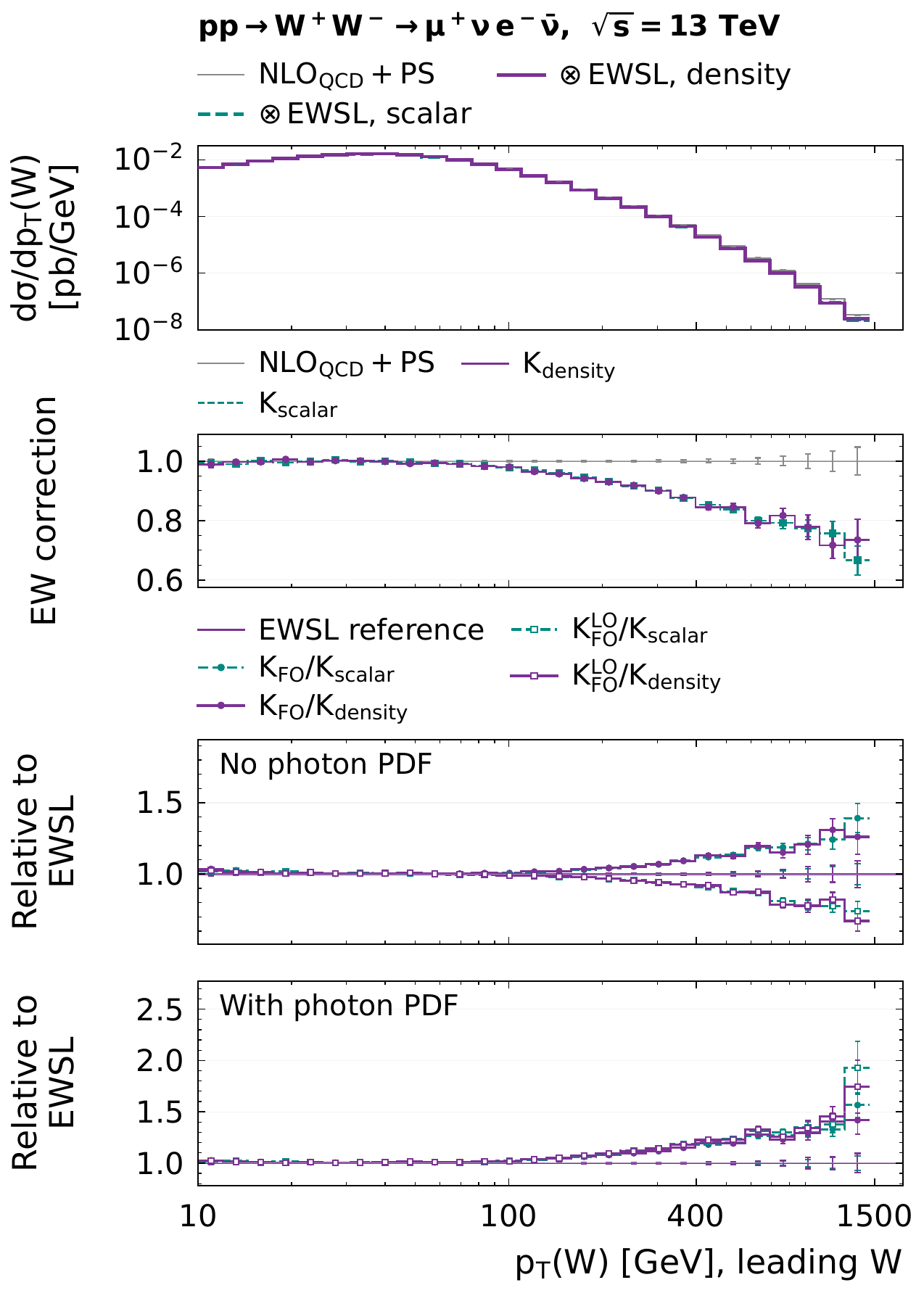}\hfill
  \includegraphics[width=0.475\textwidth]{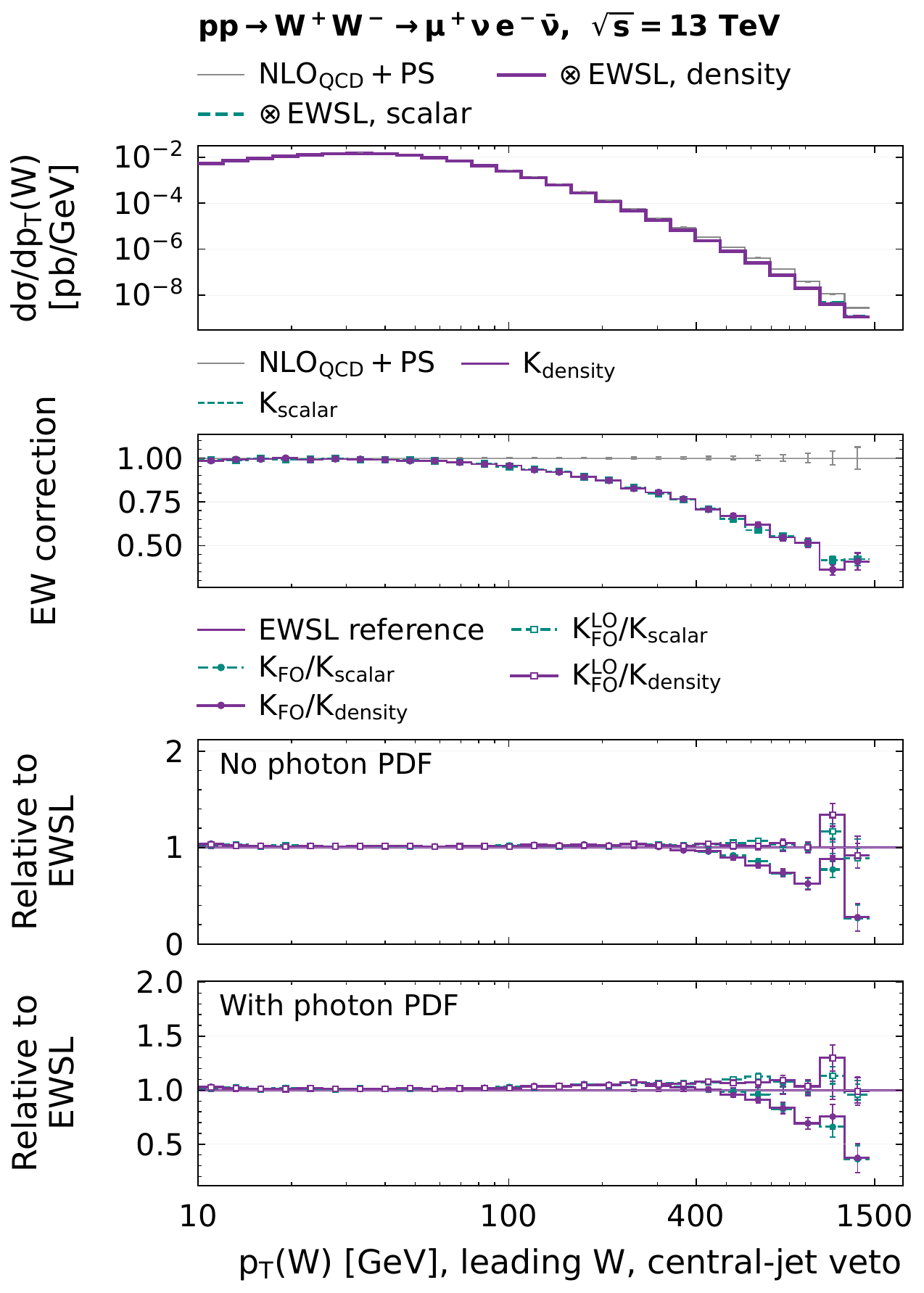}
  \caption{$W$-boson observables, defined from the Monte Carlo truth, in
    $W^+W^-$ production: $m(W^+W^-)$ (top) and the leading
    $\pT(W)$ (bottom), inclusively (left) and with a central-jet veto
    (right). Below the showered spectra are $K_{\mathrm{scalar}}$ and
    $K_{\mathrm{density}}$, then $\KFO/K_X$ and $\KFOLO/K_X$ for
    both prescriptions without and with the photon PDF. These comparison
    panels share a legend: teal dashed and purple solid denote
    scalar and density, respectively; circles use $\KFO$ and open squares $\KFOLO$.
    Error bars are statistical; boundary arrows mark truncated intervals.}
  \label{fig:ww_incl}
\end{figure}

In \cref{fig:ww_incl}, each main panel shows the unreweighted
$\NLOQCD+\mathrm{PS}$ baseline in grey, the scalar EWSL prediction in
dashed teal and the density-matrix EWSL prediction in solid purple.
The EW-correction panel shows $K_{\mathrm{scalar}}$ and
$K_{\mathrm{density}}$, with the grey baseline reference at unity.
Two panels labelled ``Relative to EWSL'' compare the fixed-order
factors with these showered corrections: the first uses zero photon
density and the second the physical photon PDF. Each shows $\KFO/K_X$
and $\KFOLO/K_X$ for $X\in\{\mathrm{scalar},\mathrm{density}\}$.  Teal
dashed and purple solid curves retain the scalar and density
conventions; filled circles and open squares distinguish the $\KFO$
and $\KFOLO$ numerators, respectively. The purple unity line, labelled
``EWSL reference'', denotes agreement with the EWSL correction. Error
bars follow the convention of \cref{fig:ewsl_validation}.  The
comparison legend is shared between the two photon-PDF panels.  Their
axis ranges resolve the central comparisons; arrowheads on the
boundaries mark statistical error bars extending outside the display.
The physical-photon-PDF panel displays the effect of the additional
partonic channels and is not a matched comparison with the
quark-initiated EWSL sample.

\Cref{fig:ww_incl} first presents the inclusive fixed-order
comparison. For $m(W^+W^-)$, the density and scalar curves overlap and
show the expected small Sudakov suppression\footnote{We remind the
  reader that even at very large $m(W^+W^-)$, not all invariants are
  necessarily large since the $W$ bosons can be quite forward.}: the
correction is close to unity near threshold, reaches a few per cent
around the TeV scale, and approaches ten per cent in the multi-TeV
tail. The no-photon fixed-order curves exhibit the same
trend. Including the photon PDF partly compensates this suppression in
the tail, illustrating why the photon-induced contribution must be
separated in a comparison with the quark-initiated EWSL weight. The
agreement of the density and scalar paths is expected for an
observable that is inclusive in the decays: after integrating over the
decay angles, the density-matrix contraction reduces to the
polarisation-summed result.

The leading-$\pT(W)$ distribution more directly probes hard transverse
recoil and consequently develops a substantially larger
correction, reaching several tens of per cent in the tail. Here the
inclusive fixed-order comparison also exposes the effect of hard QCD
radiation. Real-emission $W^+W^-+\text{jet}$ configurations can generate a
large $\pT(W)$ without making all electroweak invariants large. They
inflate the $\NLOQCD$ denominator and dilute the additive
$\KFO$, while the multiplicative EWSL
prescription also reweights the real-emission events. We therefore
repeat both distributions with a central-jet veto, rejecting
anti-$k_T$ jets with $R=0.4$, $\pT>30\GeV$, and $|\eta|<2.5$. The
veto suppresses these recoil configurations and brings the two
fixed-order normalisations and the showered EWSL correction into much
closer agreement. Together, the inclusive and vetoed results provide a
rate-level consistency check: the implementation tracks the fixed-order
weak-Sudakov behaviour once the partonic content and the effective QCD
order of the observable are matched.

We next turn to observables that depend explicitly on the decay
kinematics. Since the stable-$W$ fixed-order reference used above
cannot test these distributions, the relevant comparison is between
the density and scalar paths. To enrich the
sample in hard configurations, we require $\pT(W)>200\GeV$ for at
least one of the $W$ bosons.

\begin{figure}[htbp]
  \centering
  \includegraphics[width=0.49\textwidth]{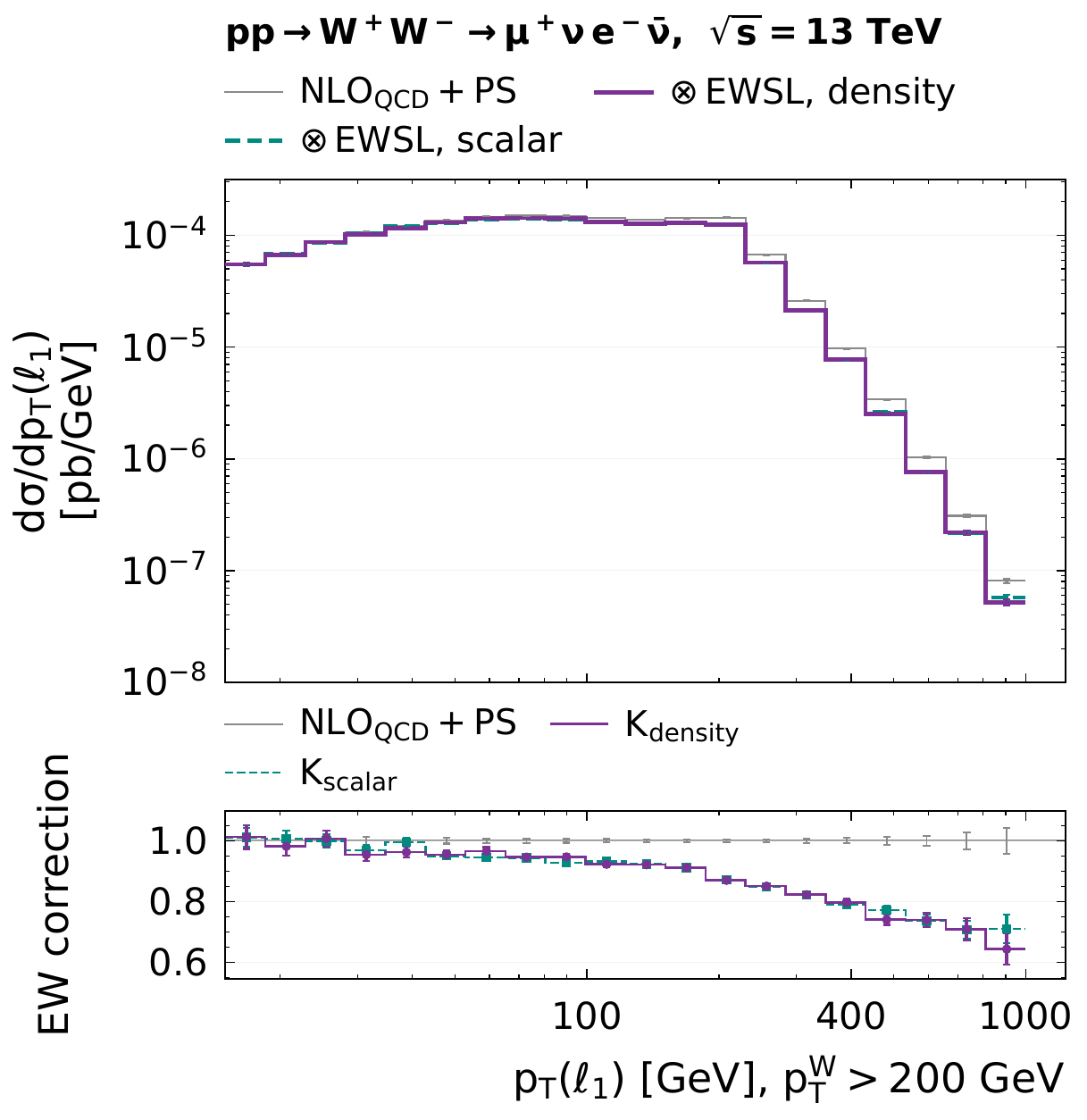}\hfill
  \includegraphics[width=0.49\textwidth]{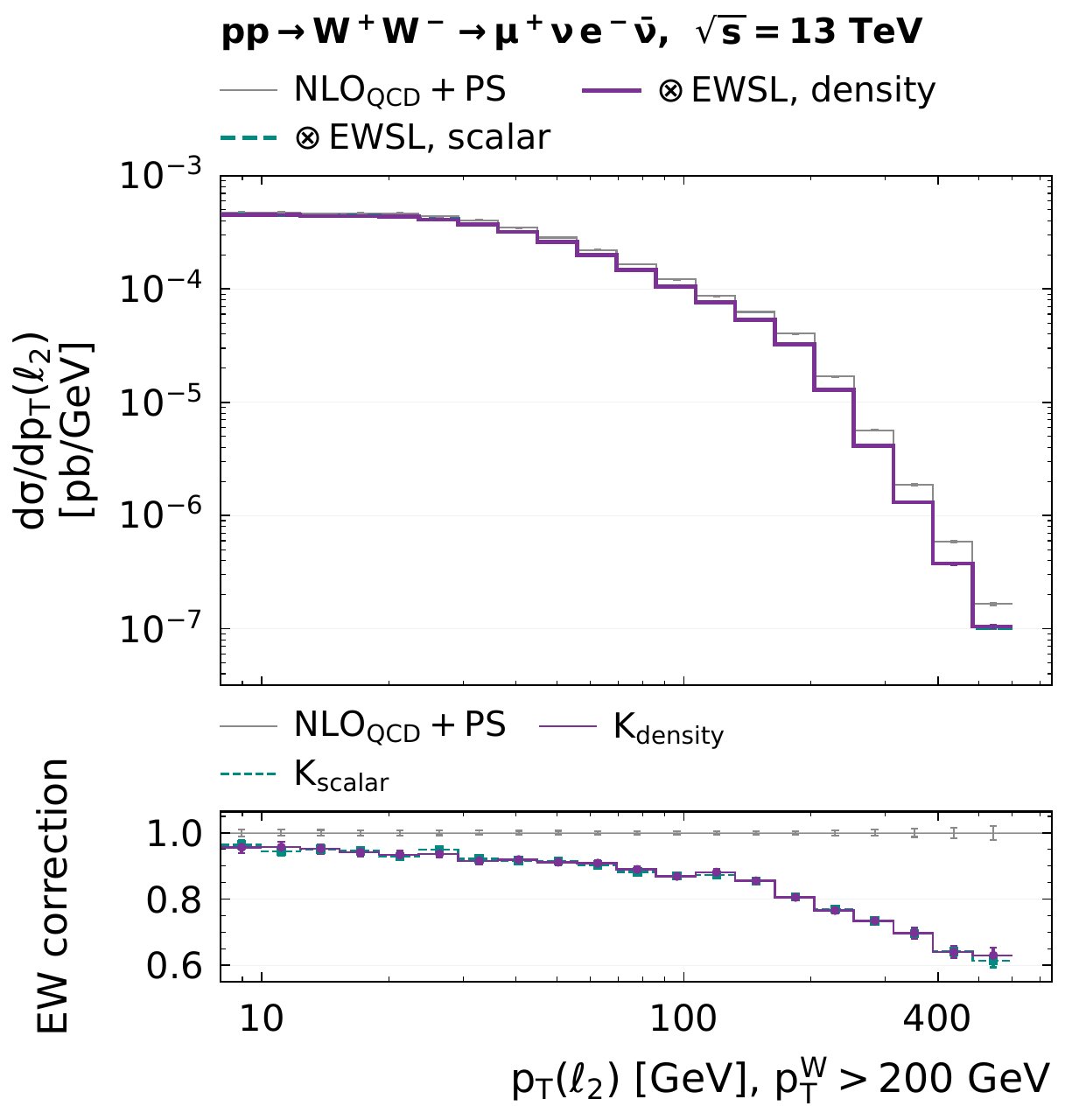}
  \caption{Transverse-momentum distributions of the two charged
    leptons for $\pT(W)>200\GeV$. The lower panels show
    $K_{\mathrm{scalar}}$ (teal dashed) and $K_{\mathrm{density}}$
    (purple solid), relative to the spin-correlated
    $\NLOQCD+\mathrm{PS}$ baseline (grey unity reference).
    Error bars are statistical.}
  \label{fig:lepton}
\end{figure}

Each main panel of \cref{fig:lepton} contains the same three showered
predictions: the baseline in grey, scalar EWSL reweighting in dashed
teal and density-matrix EWSL reweighting in solid purple. The lower
insets show $K_{\mathrm{scalar}}$ and $K_{\mathrm{density}}$, using
the same colours and line styles, with the grey unreweighted
reference at unity and statistical error bars.

The two charged-lepton spectra in \cref{fig:lepton} show a net
suppression of around ten per cent over much of the populated region,
which grows towards the high-$\pT$ tails. Relative to the QCD-only
baseline, these ratios include both the weak Sudakov weight and QED
shower migration; the suppression cannot therefore be attributed to
the Sudakov correction alone. The density and scalar results use the
same shower settings and remain statistically compatible throughout.
Since these two predictions agree for these observables, we can
conclude that for these observables there are no significant effects
due to spin correlations in the EWSL corrections.

A more direct test of spin-correlations is provided by the
signed decay-plane angle $\varphi^*$, defined as the signed angle
between the two $W$ decay planes in the $W^+W^-$ rest frame. If
spin-correlations between the two $W$ bosons were lost, this
distribution would be flat. We also use the helicity angles
$\theta_\pm$ of the charged leptons in their respective $W^\pm$ rest
frames, measured relative to the parent-$W$ flight direction in the
$W^+W^-$ rest frame, to select different regions of the decay phase
space. The correction factors $K_X$, for
$X\in\{\mathrm{scalar},\mathrm{density}\}$, retain the separate
QCD-only baseline used above, cf.~\cref{eq:nlops_correction}. For the
shape study we instead define
\begin{equation*}
  \widetilde K_X \quad \textrm{with} \quad X\in\{\mathrm{scalar},\mathrm{density}\},
\end{equation*}
where, unlike to ratio defined in \cref{eq:nlops_correction}, the
denominator also has the QED shower turned on. This makes the
numerator and denominator statistically fully correlated in our setup,
reducing statistical uncertainties in $\widetilde K_X$ and allowing
us to focus on the spin-correlation effects in EWSL
alone. Furthermore, $\langle\widetilde K_X\rangle$ is a constant
defined such that $\widetilde K_X/\langle\widetilde K_X\rangle$
becomes a normalised distribution.

\begin{figure}[htbp]
  \centering
  \includegraphics[width=0.49\textwidth]{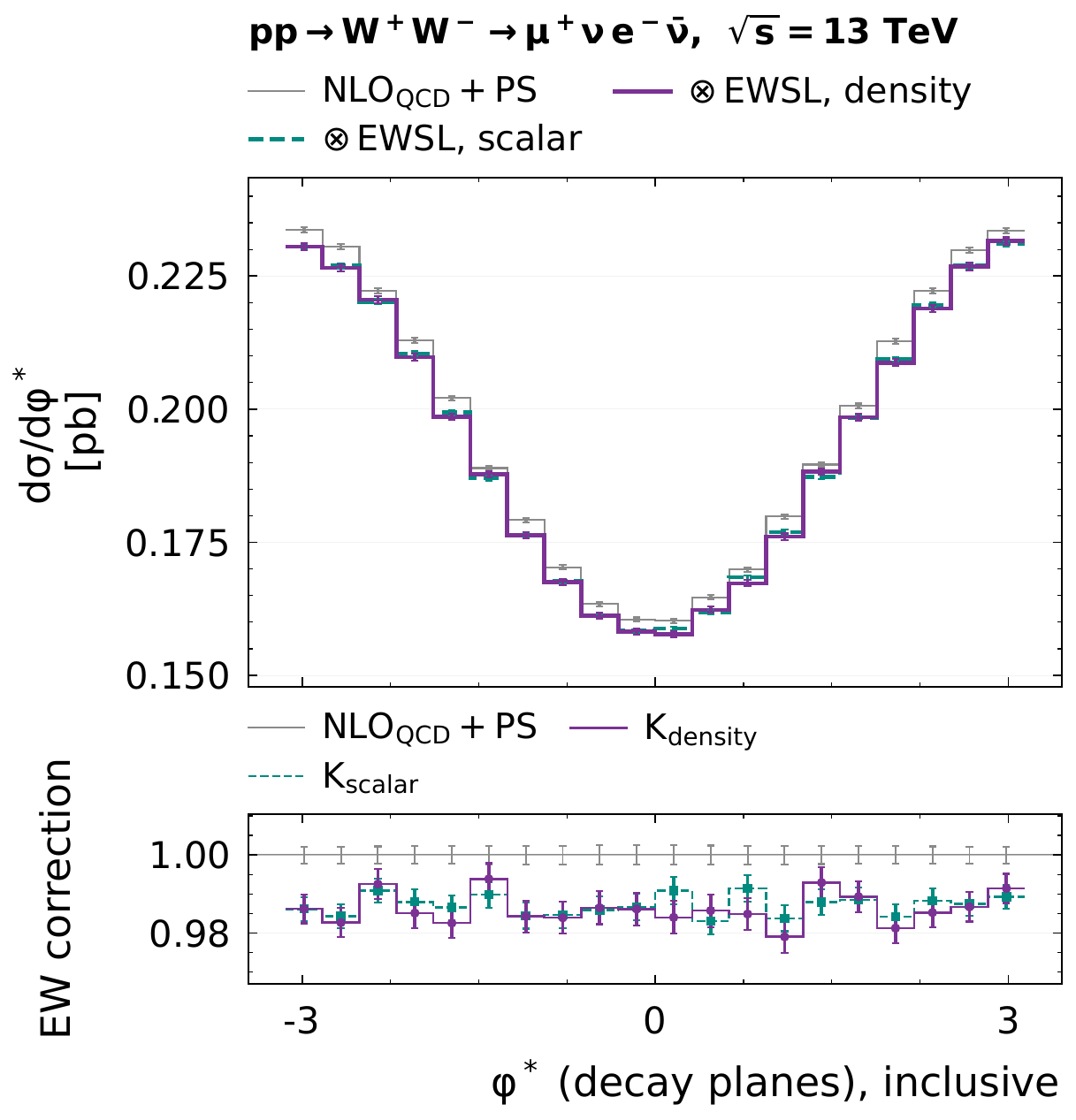}\hfill
  \includegraphics[width=0.49\textwidth]{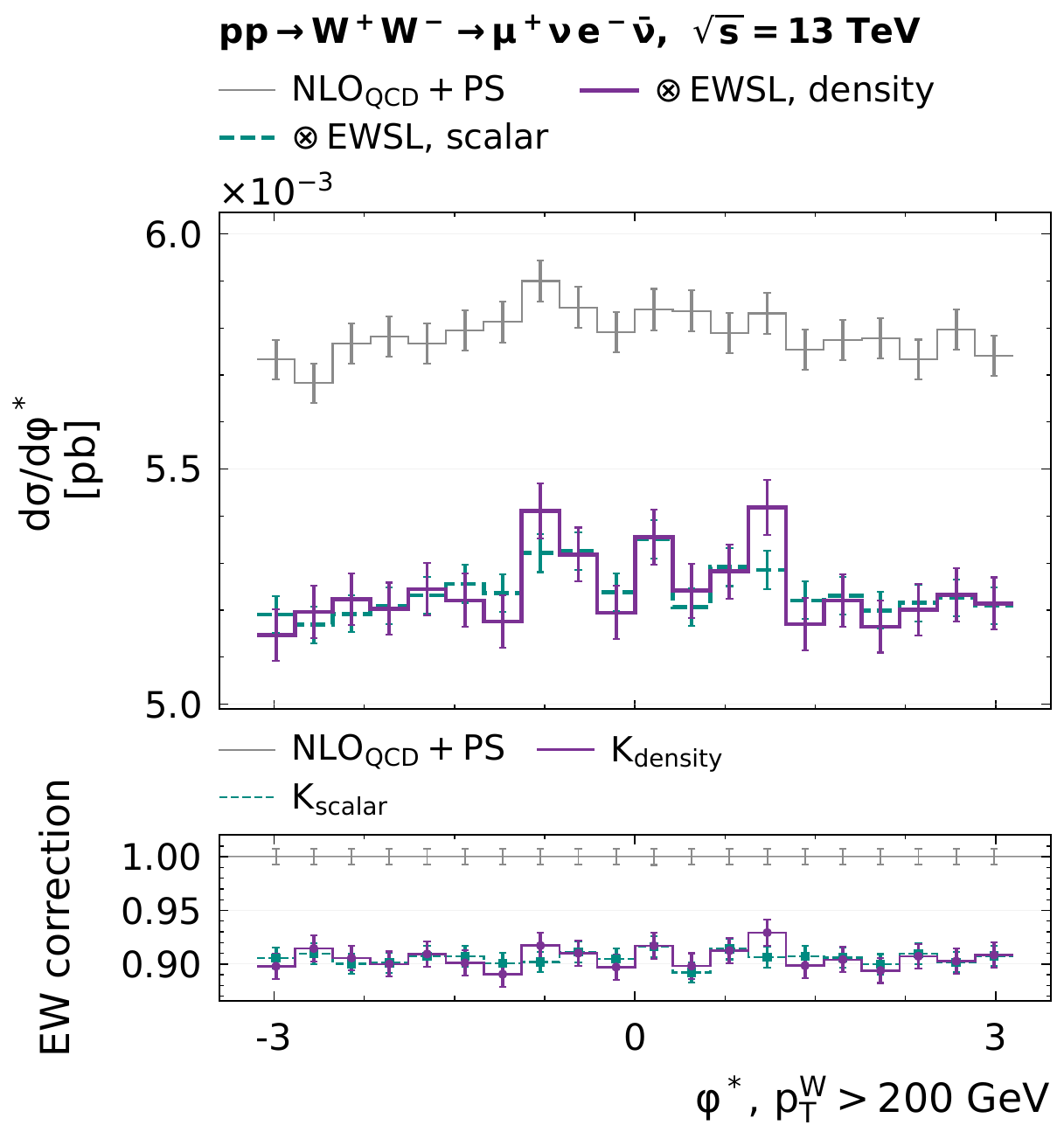}
  \caption{Spin-correlation test using the signed decay-plane angle
    $\varphi^*$. The panels show the inclusive distribution (left)
    and the distribution for $\pT(W)>200\GeV$ (right), with
    $K_{\mathrm{scalar}}$ and $K_{\mathrm{density}}$ relative to
    the spin-correlated baseline. Error bars are statistical.
    Same-sample normalised shape corrections are shown separately in \cref{fig:spin_shape}.}
  \label{fig:spin}
\end{figure}

\begin{figure}[p]
  \centering
  \includegraphics[width=0.99\textwidth]{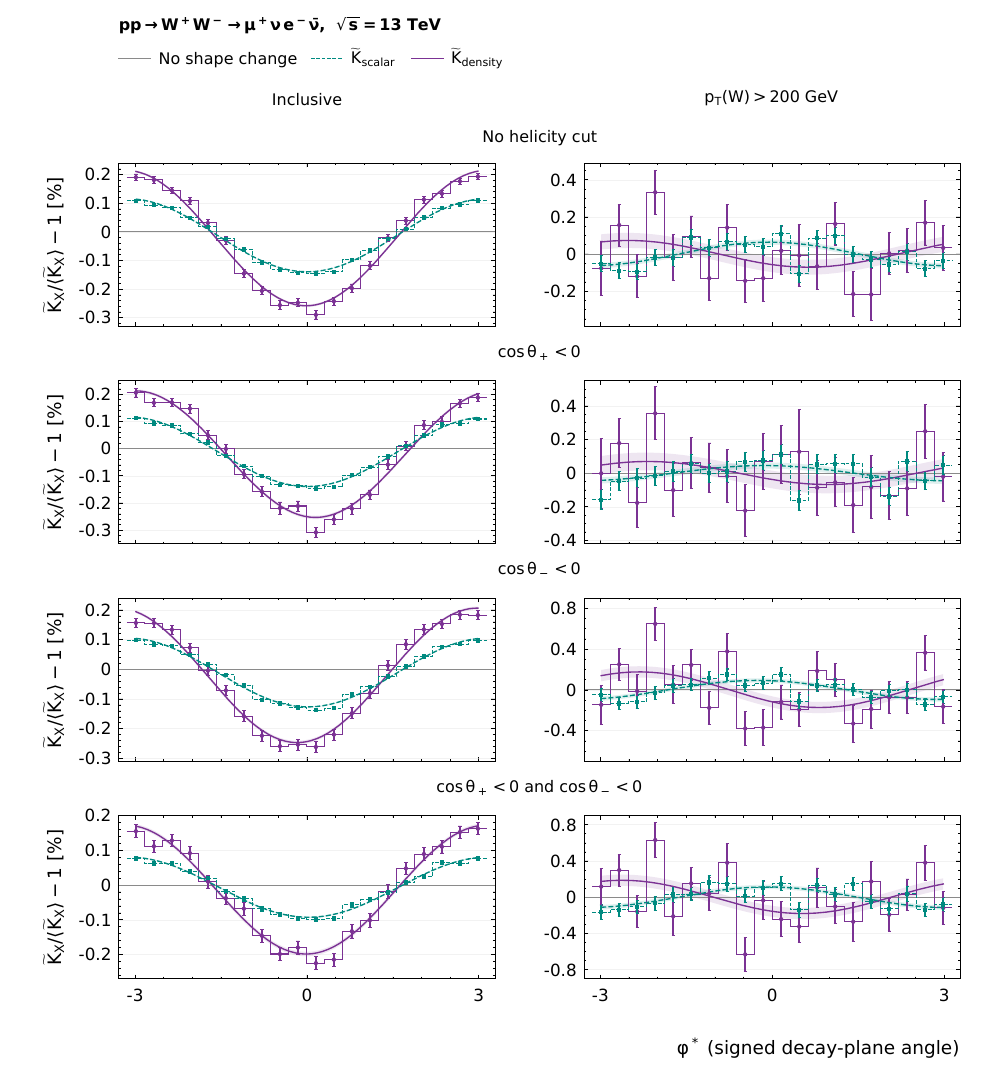}
  \caption{Residual shape dependence of the signed decay-plane angle,
    $\widetilde K_X(\varphi^*)/\langle\widetilde K_X\rangle-1$, expressed as a percentage,
    for density-matrix (purple solid) and scalar (teal dashed) reweighting.
    Each $\widetilde K_X$ uses nominal weights from the same QCD$+$QED-showered
    sample, rather than the separate QCD-only baseline of \cref{fig:spin}.
    The left column is inclusive and the right column imposes
    $\pT(W)>200\GeV$; the rows successively impose no
    helicity-angle cut, $\cos\theta_+<0$, $\cos\theta_-<0$, and both
    cuts. Steps and markers show the binned values with statistical error
    bars; the grey zero line denotes no shape change. The smooth
    lines are weighted least-squares fits of the form
    $a_0+a_1\cos\varphi^*+b_1\sin\varphi^*$, included only to guide the eye,
    with translucent bands showing their $\pm1\sigma$ envelopes. Each panel
    has its own vertical range, retaining all statistical error bars and fit
    bands. In the $\pT(W)>200\GeV$ column the band is
    comparable to the modulation itself, i.e.\ the amplitude is not
    resolved.}
  \label{fig:spin_shape}
\end{figure}

The two plots of \cref{fig:spin} use the same convention as
\cref{fig:lepton}: grey for the baseline, dashed teal for scalar EWSL
reweighting and solid purple for density-matrix EWSL reweighting.
Their lower insets show $K_{\mathrm{scalar}}$ and
$K_{\mathrm{density}}$ relative to the grey baseline reference.
The separate grid in \cref{fig:spin_shape} shows the residual shape
correction $\widetilde K_X(\varphi^*)/\langle\widetilde K_X\rangle-1$, expressed as a
percentage, with purple circles for density-matrix reweighting and
teal squares for scalar reweighting. The two columns compare the inclusive
and $\pT(W)>200\GeV$ selections; the four rows apply the different
helicity-angle selections. Smooth curves and translucent bands show
the fits and their $\pm1\sigma$ envelopes described in the caption.

The inclusive baseline in \cref{fig:spin} has a pronounced
$\varphi^*$ dependence, and both EWSL prescriptions preserve its
correlated shape. Inclusively,
the overall correction is only at the per cent level because most
events lie close to the weak scale. In the same-sample shape study, after
dividing out $\langle\widetilde K_X\rangle$, the residual $\varphi^*$ modulation in
\cref{fig:spin_shape} is only
a few per mille. The density-matrix reweighting gives a slightly larger
modulation than the scalar path, but their difference remains at the
per-mille level in the normalised distribution. The same conclusion holds
after the $\cos\theta_\pm$ selections.

The $\pT(W)>200\GeV$ cut makes the average Sudakov suppression much
larger, about ten per cent in the populated region, but does not
enhance the distortion of the normalised $\varphi^*$ shape. The hard cut
primarily increases the common logarithmic correction to the
production amplitudes. A spin-correlation shape can change only
through differences between the corrections to distinct helicity
components; a common factor cancels in
$\widetilde K_X(\varphi^*)/\langle\widetilde K_X\rangle$. In addition, the cut selects a narrower
range of hard production kinematics and substantially reduces the
event sample. Within the resulting sub-per-cent statistical precision,
no enhancement of the normalised $\varphi^*$ modulation is resolved. Thus
the cut clearly increases the rate suppression, but the present sample
does not establish a corresponding change in the spin-correlation
shape.

Taken together, the reconstructed-$W$, lepton-level, and decay-plane
comparisons provide non-trivial internal consistency checks of the
resonance-aware contraction implemented in \cref{eq:w_lambda}. They
also show that, for the $W^+W^-$ setup considered here, the scalar
approximation happens to reproduce the density result very
accurately. In fact, for all observables we have considered, including
many more than we present here in this work, we have not found
sizeable differences between the scalar and density-matrix reweighting
prescriptions.  A similar observation was made in
ref.~\cite{Lindert:2023fcu} for the two processes considered there in
the context of resonance decays, $Z+j$ and $ZZ$. In the on-shell
regions, the resonance-aware logarithmic calculation agreed well with
on-shell production followed by spin-correlated decays, indicating
small additional spin-correlation effects.  However, this numerical
agreement is specific to the processes and observables studied and
should not be interpreted as making the density construction
unnecessary for resonances with more strongly helicity-dependent
electroweak corrections. We leave further explorations of the
differences between the scalar and density-matrix reweighting
prescriptions for future work.

\FloatBarrier
\section{Conclusions}
\label{sec:conclusions}

We have extended the electroweak Sudakov reweighting of \mgamc{} in two
directions: to multijet-merged event samples and to processes containing
decaying spin-carrying resonances. The first extension reuses the
clustering history assigned by FxFx. After recorded decays have been
reversed, the history is traversed until a configuration is found whose
invariants satisfy the operational Sudakov-domain condition. The
electroweak logarithms are evaluated on this reconstructed Born-like
object rather than indiscriminately on all resolved matrix-element
partons. If no admissible configuration is found, the event is left
uncorrected.

When the reconstruction exposes an unstable resonance, we retain its
helicity indices in the production density matrix, apply the
helicity-dependent Sudakov correction there, and contract the result
with the corresponding decay matrices. This procedure propagates the
electroweak correction through the production-decay spin correlations
present in the event. It covers both decay products generated in the
hard matrix element and on-shell resonances subsequently decayed with
MadSpin. The diagonal density-matrix correction is fixed directly by
the helicity-resolved amplitudes. The off-diagonal construction uses
Born-weighted correction factors and entails an additional
approximation when these factors depend on production degrees of
freedom that have been summed over.

The FxFx reconstruction was tested in $Z+\mathrm{jets}$ production.
The reweighted prediction follows the expected high-energy
fixed-order behaviour: the correction to $\pT(Z)$ is about $-5\%$ at
$300\GeV$ and $-15\%$ at $1\TeV$, whereas the inclusive leading-jet
correction remains below ten per cent because hard dijet recoil need not
correspond to a hard electroweak configuration. A second-jet veto
restores close agreement with the LO-normalised fixed-order correction.
For the distributions studied, varying $Q_{\mathrm{cut}}$ between
$20$ and $40\GeV$ produces the usual few-per-cent FxFx spread in the
tails, while varying the clustering-threshold parameter
$\xi_{\mathrm{clust}}$ between $1$ and $2$
changes the electroweak weight only at the sub-per-cent level.

The density-matrix reweighting was tested in fully leptonic $W^+W^-$
production. For observables inclusive in the decays, the density and
scalar prescriptions agree, as expected after integration over the
decay angles. The weak correction approaches ten per cent in the
multi-TeV $W^+W^-$ invariant-mass tail and several tens of per cent in
the leading-$W$ transverse-momentum tail. Lepton and decay-plane
observables probe the reweighting of production-decay spin
correlations. For this channel, the
additional helicity dependence of the Sudakov correction is small: the
density and scalar predictions are statistically compatible, and
their difference in normalised $\varphi^*$ distributions is at the
per-mille level.

The result is a QED-subtracted virtual weak-Sudakov approximation, not
an exact NLO electroweak prediction. It omits non-logarithmic
one-loop terms, the complete fixed-order QED contribution, and
photon-initiated channels not already present
in the underlying sample. Its multiplicative application also
generates approximate mixed QCD-electroweak contributions beyond the
orders controlled by the hard events. The present numerical tests
cover a stable $Z$ and on-shell $W$ bosons in the narrow-width
approximation. Comparisons with exact decay-level NLO-EW calculations,
and tests of hard-matrix-element decays, top-quark resonances, and
channels with more strongly helicity-dependent weak corrections are
therefore natural next steps.

\section*{Acknowledgements}
L.J.~and M.M.~have been supported by Slovenian Research and Innovation
Agency grants J1-50217, N1-0398 and P1-0135.
D.P.~and M.Z.~acknowledge the financial support from the MUR (Italy),
with funds of the European Union (NextGenerationEU), through the
PRIN2022 grant 2022EZ3S3F.

\bibliographystyle{unsrt}
\bibliography{references}

@article{Actis:2012qn,
    author = "Actis, S. and Denner, A. and Hofer, L. and Scharf, A. and Uccirati, S.",
    title = "{Recursive generation of one-loop amplitudes in the Standard Model}",
    eprint = "1211.6316",
    archivePrefix = "arXiv",
    primaryClass = "hep-ph",
    reportNumber = "PSI-PR-12-09",
    doi = "10.1007/JHEP04(2013)037",
    journal = "JHEP",
    volume = "04",
    pages = "037",
    year = "2013"
}

@article{Actis:2016mpe,
    author = "Actis, Stefano and Denner, Ansgar and Hofer, Lars and Lang, Jean-Nicolas and Scharf, Andreas and Uccirati, Sandro",
    title = "{RECOLA: REcursive Computation of One-Loop Amplitudes}",
    eprint = "1605.01090",
    archivePrefix = "arXiv",
    primaryClass = "hep-ph",
    reportNumber = "ICCUB-16-020",
    doi = "10.1016/j.cpc.2017.01.004",
    journal = "Comput. Phys. Commun.",
    volume = "214",
    pages = "140--173",
    year = "2017"
}

@article{Alioli:2010xd,
    author = "Alioli, Simone and Nason, Paolo and Oleari, Carlo and Re, Emanuele",
    title = "{A general framework for implementing NLO calculations in shower Monte Carlo programs: the POWHEG BOX}",
    eprint = "1002.2581",
    archivePrefix = "arXiv",
    primaryClass = "hep-ph",
    reportNumber = "DESY-10-018, SFB-CPP-10-22, IPPP-10-11, DCPT-10-22",
    doi = "10.1007/JHEP06(2010)043",
    journal = "JHEP",
    volume = "06",
    pages = "043",
    year = "2010"
}

@article{Alwall:2014hca,
    author = "Alwall, J. and Frederix, R. and Frixione, S. and Hirschi, V. and Maltoni, F. and Mattelaer, O. and Shao, H. -S. and Stelzer, T. and Torrielli, P. and Zaro, M.",
    title = "{The automated computation of tree-level and next-to-leading order differential cross sections, and their matching to parton shower simulations}",
    eprint = "1405.0301",
    archivePrefix = "arXiv",
    primaryClass = "hep-ph",
    reportNumber = "CERN-PH-TH-2014-064, CP3-14-18, LPN14-066, MCNET-14-09, ZU-TH-14-14",
    doi = "10.1007/JHEP07(2014)079",
    journal = "JHEP",
    volume = "07",
    pages = "079",
    year = "2014"
}

@article{Artoisenet:2012st,
    author = "Artoisenet, Pierre and Frederix, Rikkert and Mattelaer, Olivier and Rietkerk, Robbert",
    title = "{Automatic spin-entangled decays of heavy resonances in Monte Carlo simulations}",
    eprint = "1212.3460",
    archivePrefix = "arXiv",
    primaryClass = "hep-ph",
    reportNumber = "NIKHEF-2012-021, CERN-PH-TH-2012-329",
    doi = "10.1007/JHEP03(2013)015",
    journal = "JHEP",
    volume = "03",
    pages = "015",
    year = "2013"
}

@article{Barze:2012tt,
    author = "Barze, Luca and Montagna, Guido and Nason, Paolo and Nicrosini, Oreste and Piccinini, Fulvio",
    title = "{Implementation of electroweak corrections in the POWHEG BOX: single W production}",
    eprint = "1202.0465",
    archivePrefix = "arXiv",
    primaryClass = "hep-ph",
    reportNumber = "CERN-PH-TH-2012-025, FNT-2012-01, LPN12-031",
    doi = "10.1007/JHEP04(2012)037",
    journal = "JHEP",
    volume = "04",
    pages = "037",
    year = "2012"
}

@article{Barze:2013fru,
    author = "Barze, Luca and Montagna, Guido and Nason, Paolo and Nicrosini, Oreste and Piccinini, Fulvio and Vicini, Alessandro",
    title = "{Neutral current Drell-Yan with combined QCD and electroweak corrections in the POWHEG BOX}",
    eprint = "1302.4606",
    archivePrefix = "arXiv",
    primaryClass = "hep-ph",
    reportNumber = "CERN-PH-TH-2013-027, FNT-2013-01, IFUM-1007-FT, LPN13-018",
    doi = "10.1140/epjc/s10052-013-2474-y",
    journal = "Eur. Phys. J. C",
    volume = "73",
    number = "6",
    pages = "2474",
    year = "2013"
}

@misc{Bellm:2025pcw,
    author = "Bellm, J. and others",
    title = "{The Physics of Herwig 7}",
    eprint = "2512.16645",
    archivePrefix = "arXiv",
    primaryClass = "hep-ph",
    reportNumber = "CERN-TH-2025-252, IPPP-25-57, HERWIG-2025-01, KA-TP-36-2025, MCNET-25-31",
    howpublished = "{arXiv:2512.16645 [hep-ph]}",
    month = "12",
    year = "2025"
}

@article{Bertone:2017bme,
    author = "Bertone, Valerio and Carrazza, Stefano and Hartland, Nathan P. and Rojo, Juan",
    collaboration = "NNPDF",
    title = "{Illuminating the photon content of the proton within a global PDF analysis}",
    eprint = "1712.07053",
    archivePrefix = "arXiv",
    primaryClass = "hep-ph",
    reportNumber = "Nikhef/2017-064, CERN-TH-2017-235, NIKHEF-2017-064",
    doi = "10.21468/SciPostPhys.5.1.008",
    journal = "SciPost Phys.",
    volume = "5",
    number = "1",
    pages = "008",
    year = "2018"
}

@article{Biedermann:2017yoi,
    author = {Biedermann, Benedikt and Br{\"a}uer, Stephan and Denner, Ansgar and Pellen, Mathieu and Schumann, Steffen and Thompson, Jennifer M.},
    title = "{Automation of NLO QCD and EW corrections with Sherpa and Recola}",
    eprint = "1704.05783",
    archivePrefix = "arXiv",
    primaryClass = "hep-ph",
    doi = "10.1140/epjc/s10052-017-5054-8",
    journal = "Eur. Phys. J. C",
    volume = "77",
    pages = "492",
    year = "2017"
}

@article{Bierlich:2022pfr,
    author = "Bierlich, Christian and others",
    title = "{A comprehensive guide to the physics and usage of PYTHIA 8.3}",
    eprint = "2203.11601",
    archivePrefix = "arXiv",
    primaryClass = "hep-ph",
    reportNumber = "LU-TP 22-16, MCNET-22-04, FERMILAB-PUB-22-227-SCD",
    doi = "10.21468/SciPostPhysCodeb.8",
    journal = "SciPost Phys. Codeb.",
    volume = "2022",
    pages = "8",
    year = "2022"
}

@article{Bothmann:2020sxm,
    author = "Bothmann, Enrico and Napoletano, Davide",
    title = "{Automated evaluation of electroweak Sudakov logarithms in Sherpa}",
    eprint = "2006.14635",
    archivePrefix = "arXiv",
    primaryClass = "hep-ph",
    doi = "10.1140/epjc/s10052-020-08596-2",
    journal = "Eur. Phys. J. C",
    volume = "80",
    number = "11",
    pages = "1024",
    year = "2020"
}

@article{Brauer:2020kfv,
    author = {Br{\"a}uer, Stephan and Denner, Ansgar and Pellen, Mathieu and Sch{\"o}nherr, Marek and Schumann, Steffen},
    title = "{Fixed-order and merged parton-shower predictions for WW and WWj production at the LHC including NLO QCD and EW corrections}",
    eprint = "2005.12128",
    archivePrefix = "arXiv",
    primaryClass = "hep-ph",
    reportNumber = "Cavendish-HEP 20/05, IPPP/20/19, MCNET-20-14",
    doi = "10.1007/JHEP10(2020)159",
    journal = "JHEP",
    volume = "10",
    pages = "159",
    year = "2020"
}

@article{Buccioni:2019sur,
    author = {Buccioni, Federico and Lang, Jean-Nicolas and Lindert, Jonas M. and Maierh{\"o}fer, Philipp and Pozzorini, Stefano and Zhang, Hantian and Zoller, Max F.},
    title = "{OpenLoops 2}",
    eprint = "1907.13071",
    archivePrefix = "arXiv",
    primaryClass = "hep-ph",
    reportNumber = "IPPP/19/62, FR-PHENO-2019-12, PSI-PR-19-15, ZU-TH 37/19",
    doi = "10.1140/epjc/s10052-019-7306-2",
    journal = "Eur. Phys. J. C",
    volume = "79",
    number = "10",
    pages = "866",
    year = "2019"
}

@article{Cascioli:2011va,
    author = "Cascioli, Fabio and Maierhofer, Philipp and Pozzorini, Stefano",
    title = "{Scattering Amplitudes with Open Loops}",
    eprint = "1111.5206",
    archivePrefix = "arXiv",
    primaryClass = "hep-ph",
    reportNumber = "ZU-TH-23-11, LPN11-66",
    doi = "10.1103/PhysRevLett.108.111601",
    journal = "Phys. Rev. Lett.",
    volume = "108",
    pages = "111601",
    year = "2012"
}

@article{Chiesa:2015mya,
    author = "Chiesa, Mauro and Greiner, Nicolas and Tramontano, Francesco",
    title = "{Automation of electroweak corrections for LHC processes}",
    eprint = "1507.08579",
    archivePrefix = "arXiv",
    primaryClass = "hep-ph",
    reportNumber = "DESY-15-113, FNT-2015-01",
    doi = "10.1088/0954-3899/43/1/013002",
    journal = "J. Phys. G",
    volume = "43",
    number = "1",
    pages = "013002",
    year = "2016"
}

@article{Chiesa:2017gqx,
    author = {Chiesa, Mauro and Greiner, Nicolas and Sch{\"o}nherr, Marek and Tramontano, Francesco},
    title = "{Electroweak corrections to diphoton plus jets}",
    eprint = "1706.09022",
    archivePrefix = "arXiv",
    primaryClass = "hep-ph",
    reportNumber = "ZU--TH-20-17-MCNET--17--11",
    doi = "10.1007/JHEP10(2017)181",
    journal = "JHEP",
    volume = "10",
    pages = "181",
    year = "2017"
}

@article{Chiesa:2019ulk,
    author = "Chiesa, Mauro and Denner, Ansgar and Lang, Jean-Nicolas and Pellen, Mathieu",
    title = "{An event generator for same-sign W-boson scattering at the LHC including electroweak corrections}",
    eprint = "1906.01863",
    archivePrefix = "arXiv",
    primaryClass = "hep-ph",
    reportNumber = "Cavendish-HEP-19/09, ZU-TH-28/19, VBSCAN-PUB-04-19",
    doi = "10.1140/epjc/s10052-019-7290-6",
    journal = "Eur. Phys. J. C",
    volume = "79",
    number = "9",
    pages = "788",
    year = "2019"
}

@article{Chiesa:2020ttl,
    author = "Chiesa, Mauro and Oleari, Carlo and Re, Emanuele",
    title = "{NLO QCD+NLO EW corrections to diboson production matched to parton shower}",
    eprint = "2005.12146",
    archivePrefix = "arXiv",
    primaryClass = "hep-ph",
    reportNumber = "LAPTH-025/20",
    doi = "10.1140/epjc/s10052-020-8419-3",
    journal = "Eur. Phys. J. C",
    volume = "80",
    number = "9",
    pages = "849",
    year = "2020"
}

@article{Cullen:2011ac,
    author = "Cullen, Gavin and Greiner, Nicolas and Heinrich, Gudrun and Luisoni, Gionata and Mastrolia, Pierpaolo and Ossola, Giovanni and Reiter, Thomas and Tramontano, Francesco",
    collaboration = "GoSam",
    title = "{Automated One-Loop Calculations with GoSam}",
    eprint = "1111.2034",
    archivePrefix = "arXiv",
    primaryClass = "hep-ph",
    reportNumber = "CERN-PH-TH-2011-272, MPP-2011-132, DESY-11-203, IPPP-11-72, DCPT-11-144, EDINBURGH-2011-32",
    doi = "10.1140/epjc/s10052-012-1889-1",
    journal = "Eur. Phys. J. C",
    volume = "72",
    pages = "1889",
    year = "2012"
}

@article{Denner:2000jv,
    author = "Denner, Ansgar and Pozzorini, Stefano",
    title = "{One loop leading logarithms in electroweak radiative corrections. 1. Results}",
    eprint = "hep-ph/0010201",
    archivePrefix = "arXiv",
    reportNumber = "PSI-PR-00-15, ZU-TH-17-00",
    doi = "10.1007/s100520100551",
    journal = "Eur. Phys. J. C",
    volume = "18",
    pages = "461--480",
    year = "2001"
}

@article{Denner:2001gw,
    author = "Denner, Ansgar and Pozzorini, Stefano",
    title = "{One loop leading logarithms in electroweak radiative corrections. 2. Factorization of collinear singularities}",
    eprint = "hep-ph/0104127",
    archivePrefix = "arXiv",
    reportNumber = "PSI-PR-01-06, ZU-TH-13-01",
    doi = "10.1007/s100520100721",
    journal = "Eur. Phys. J. C",
    volume = "21",
    pages = "63--79",
    year = "2001"
}

@article{Denner:2003wi,
    author = "Denner, Ansgar and Melles, M. and Pozzorini, S.",
    title = "{Two loop electroweak angular dependent logarithms at high-energies}",
    eprint = "hep-ph/0301241",
    archivePrefix = "arXiv",
    reportNumber = "PSI-PR-03-01, TTP-03-03",
    doi = "10.1016/S0550-3213(03)00307-9",
    journal = "Nucl. Phys. B",
    volume = "662",
    pages = "299--333",
    year = "2003"
}

@article{Denner:2004iz,
    author = "Denner, Ansgar and Pozzorini, S.",
    title = "{An Algorithm for the high-energy expansion of multi-loop diagrams to next-to-leading logarithmic accuracy}",
    eprint = "hep-ph/0408068",
    archivePrefix = "arXiv",
    reportNumber = "SFB-CPP-04-31, TTP04-17, PSI-PR-04-09",
    doi = "10.1016/j.nuclphysb.2005.03.036",
    journal = "Nucl. Phys. B",
    volume = "717",
    pages = "48--85",
    year = "2005"
}

@article{Denner:2006jr,
    author = "Denner, Ansgar and Jantzen, B. and Pozzorini, S.",
    title = "{Two-loop electroweak next-to-leading logarithmic corrections to massless fermionic processes}",
    eprint = "hep-ph/0608326",
    archivePrefix = "arXiv",
    reportNumber = "PSI-PR-06-10, MPP-2006-109",
    doi = "10.1016/j.nuclphysb.2006.10.014",
    journal = "Nucl. Phys. B",
    volume = "761",
    pages = "1--62",
    year = "2007"
}

@article{Denner:2008yn,
    author = "Denner, Ansgar and Jantzen, Bernd and Pozzorini, Stefano",
    title = "{Two-loop electroweak next-to-leading logarithms for processes involving heavy quarks}",
    eprint = "0809.0800",
    archivePrefix = "arXiv",
    primaryClass = "hep-ph",
    reportNumber = "PSI-PR-08-11, MPP-2008-89",
    doi = "10.1088/1126-6708/2008/11/062",
    journal = "JHEP",
    volume = "11",
    pages = "062",
    year = "2008"
}

@article{Denner:2017wsf,
    author = "Denner, Ansgar and Lang, Jean-Nicolas and Uccirati, Sandro",
    title = "{Recola2: REcursive Computation of One-Loop Amplitudes 2}",
    eprint = "1711.07388",
    archivePrefix = "arXiv",
    primaryClass = "hep-ph",
    doi = "10.1016/j.cpc.2017.11.013",
    journal = "Comput. Phys. Commun.",
    volume = "224",
    pages = "346--361",
    year = "2018"
}

@article{Durupt:2025wuk,
    author = "Durupt, Valentin and Maltoni, Fabio and Mattelaer, Olivier",
    title = "{Automated computation of spin-density matrices and quantum observables for collider physics}",
    eprint = "2510.17730",
    archivePrefix = "arXiv",
    primaryClass = "hep-ph",
    reportNumber = "IRMP-CP3-25-34",
    doi = "10.1007/JHEP04(2026)103",
    journal = "JHEP",
    volume = "04",
    pages = "103",
    year = "2026"
}

@misc{Flower:2026byh,
    author = {Flower, Lois and Sch{\"o}nherr, Marek},
    title = "{$e^+e^- \to ZH$ at NLO EW matched to a QED parton shower}",
    eprint = "2603.05585",
    archivePrefix = "arXiv",
    primaryClass = "hep-ph",
    reportNumber = "IPPP/26/24, MCnet-26-03",
    howpublished = "{arXiv:2603.05585 [hep-ph]}",
    month = "3",
    year = "2026"
}

@misc{Flower:2026lui,
    author = {Flower, Lois and Roper, Joanne and Sch{\"o}nherr, Marek},
    title = "{A resonance-aware MC@NLO QCD+EW-matched calculation of lepton-pair production}",
    eprint = "2607.06059",
    archivePrefix = "arXiv",
    primaryClass = "hep-ph",
    reportNumber = "IPPP/26/55, MCnet-26-21",
    howpublished = "{arXiv:2607.06059 [hep-ph]}",
    month = "7",
    year = "2026"
}

@article{Frederix:2009yq,
    author = "Frederix, Rikkert and Frixione, Stefano and Maltoni, Fabio and Stelzer, Tim",
    title = "{Automation of next-to-leading order computations in QCD: The FKS subtraction}",
    eprint = "0908.4272",
    archivePrefix = "arXiv",
    primaryClass = "hep-ph",
    reportNumber = "CERN-TH-2009-157, CP3-09-39",
    doi = "10.1088/1126-6708/2009/10/003",
    journal = "JHEP",
    volume = "10",
    pages = "003",
    year = "2009"
}

@article{Frederix:2012ps,
    author = "Frederix, Rikkert and Frixione, Stefano",
    title = "{Merging meets matching in MC@NLO}",
    eprint = "1209.6215",
    archivePrefix = "arXiv",
    primaryClass = "hep-ph",
    reportNumber = "CERN-PH-TH-2012-247, ZU-TH-21-12",
    doi = "10.1007/JHEP12(2012)061",
    journal = "JHEP",
    volume = "12",
    pages = "061",
    year = "2012"
}

@article{Frederix:2018nkq,
    author = "Frederix, R. and Frixione, S. and Hirschi, V. and Pagani, D. and Shao, H. -S. and Zaro, M.",
    title = "{The automation of next-to-leading order electroweak calculations}",
    eprint = "1804.10017",
    archivePrefix = "arXiv",
    primaryClass = "hep-ph",
    reportNumber = "Nikhef/2018-015, TUM-HEP-1138/18, NIKHEF-2018-015, TUM-HEP-1138-18",
    doi = "10.1007/JHEP11(2021)085",
    journal = "JHEP",
    volume = "07",
    pages = "185",
    year = "2018",
    note = "[Erratum: JHEP 11, 085 (2021)]"
}

@article{Frederix:2021agh,
    author = "Frederix, Rikkert and Tsinikos, Ioannis",
    title = "{On improving NLO merging for $ \mathrm{t}\overline{\mathrm{t}}\mathrm{W} $ production}",
    eprint = "2108.07826",
    archivePrefix = "arXiv",
    primaryClass = "hep-ph",
    reportNumber = "LU-TP 21-34",
    doi = "10.1007/JHEP11(2021)029",
    journal = "JHEP",
    volume = "11",
    pages = "029",
    year = "2021"
}

@article{Frixione:1995ms,
    author = "Frixione, S. and Kunszt, Z. and Signer, A.",
    title = "{Three jet cross-sections to next-to-leading order}",
    eprint = "hep-ph/9512328",
    archivePrefix = "arXiv",
    reportNumber = "SLAC-PUB-7073, SLAC-PUB-95-7073, ETH-TH-95-42",
    doi = "10.1016/0550-3213(96)00110-1",
    journal = "Nucl. Phys. B",
    volume = "467",
    pages = "399--442",
    year = "1996"
}

@article{Frixione:2002ik,
    author = "Frixione, Stefano and Webber, Bryan R.",
    title = "{Matching NLO QCD computations and parton shower simulations}",
    eprint = "hep-ph/0204244",
    archivePrefix = "arXiv",
    reportNumber = "CAVENDISH-HEP-02-01, LAPTH-905-02, GEF-TH-2-2002",
    doi = "10.1088/1126-6708/2002/06/029",
    journal = "JHEP",
    volume = "06",
    pages = "029",
    year = "2002"
}

@article{Frixione:2007vw,
    author = "Frixione, Stefano and Nason, Paolo and Oleari, Carlo",
    title = "{Matching NLO QCD computations with Parton Shower simulations: the POWHEG method}",
    eprint = "0709.2092",
    archivePrefix = "arXiv",
    primaryClass = "hep-ph",
    reportNumber = "BICOCCA-FT-07-9, GEF-TH-21-2007",
    doi = "10.1088/1126-6708/2007/11/070",
    journal = "JHEP",
    volume = "11",
    pages = "070",
    year = "2007"
}

@article{Frixione:2015zaa,
    author = "Frixione, S. and Hirschi, V. and Pagani, D. and Shao, H. -S. and Zaro, M.",
    title = "{Electroweak and QCD corrections to top-pair hadroproduction in association with heavy bosons}",
    eprint = "1504.03446",
    archivePrefix = "arXiv",
    primaryClass = "hep-ph",
    reportNumber = "CERN-PH-TH-2015-083, CP3-15-10, LPN15-022, SLAC-PUB-16253",
    doi = "10.1007/JHEP06(2015)184",
    journal = "JHEP",
    volume = "06",
    pages = "184",
    year = "2015"
}

@article{Granata:2017iod,
    author = "Granata, Federico and Lindert, Jonas M. and Oleari, Carlo and Pozzorini, Stefano",
    title = "{NLO QCD+EW predictions for HV and HV +jet production including parton-shower effects}",
    eprint = "1706.03522",
    archivePrefix = "arXiv",
    primaryClass = "hep-ph",
    doi = "10.1007/JHEP09(2017)012",
    journal = "JHEP",
    volume = "09",
    pages = "012",
    year = "2017"
}

@article{Gutschow:2018tuk,
    author = {G{\"u}tschow, Christian and Lindert, Jonas M. and Sch{\"o}nherr, Marek},
    title = "{Multi-jet merged top-pair production including electroweak corrections}",
    eprint = "1803.00950",
    archivePrefix = "arXiv",
    primaryClass = "hep-ph",
    reportNumber = "CERN-TH-18-039, CERN-TH-2018-039, IPPP/18/13, MCnet-18-05, IPPP-18-13, MCNET-18-05",
    doi = "10.1140/epjc/s10052-018-5804-2",
    journal = "Eur. Phys. J. C",
    volume = "78",
    number = "4",
    pages = "317",
    year = "2018"
}

@article{Hamilton:2012np,
    author = "Hamilton, Keith and Nason, Paolo and Zanderighi, Giulia",
    title = "{MINLO: Multi-Scale Improved NLO}",
    eprint = "1206.3572",
    archivePrefix = "arXiv",
    primaryClass = "hep-ph",
    reportNumber = "CERN-PH-TH-2012-166, OUTP-12-11P, MCNET-12-07",
    doi = "10.1007/JHEP10(2012)155",
    journal = "JHEP",
    volume = "10",
    pages = "155",
    year = "2012"
}

@article{Hirschi:2011pa,
    author = "Hirschi, Valentin and Frederix, Rikkert and Frixione, Stefano and Garzelli, Maria Vittoria and Maltoni, Fabio and Pittau, Roberto",
    title = "{Automation of one-loop QCD corrections}",
    eprint = "1103.0621",
    archivePrefix = "arXiv",
    primaryClass = "hep-ph",
    reportNumber = "CERN-PH-TH-2011-031, CP3-11-07, ZU-TH-01-11",
    doi = "10.1007/JHEP05(2011)044",
    journal = "JHEP",
    volume = "05",
    pages = "044",
    year = "2011"
}

@article{Hoeche:2012yf,
    author = "Hoeche, Stefan and Krauss, Frank and Schonherr, Marek and Siegert, Frank",
    title = "{QCD matrix elements + parton showers: The NLO case}",
    eprint = "1207.5030",
    archivePrefix = "arXiv",
    primaryClass = "hep-ph",
    reportNumber = "SLAC-PUB-15191, IPPP-12-52, DCPT-12-104, LPN12-081, MCNET-12-09, FR-PHENO-2012-017",
    doi = "10.1007/JHEP04(2013)027",
    journal = "JHEP",
    volume = "04",
    pages = "027",
    year = "2013"
}

@article{Kallweit:2014xda,
    author = {Kallweit, Stefan and Lindert, Jonas M. and Maierh{\"o}fer, Philipp and Pozzorini, Stefano and Sch{\"o}nherr, Marek},
    title = "{NLO electroweak automation and precise predictions for W+multijet production at the LHC}",
    eprint = "1412.5157",
    archivePrefix = "arXiv",
    primaryClass = "hep-ph",
    reportNumber = "LPN14-127, IPPP-14-107, DCPT-14-214, MCNET-14-26, ZU-TH-42-14, MITP-14-102",
    doi = "10.1007/JHEP04(2015)012",
    journal = "JHEP",
    volume = "04",
    pages = "012",
    year = "2015"
}

@article{Kallweit:2015dum,
    author = {Kallweit, Stefan and Lindert, Jonas M. and Maierhofer, Philipp and Pozzorini, Stefano and Sch{\"o}nherr, Marek},
    title = "{NLO QCD+EW predictions for V + jets including off-shell vector-boson decays and multijet merging}",
    eprint = "1511.08692",
    archivePrefix = "arXiv",
    primaryClass = "hep-ph",
    reportNumber = "DCPT-15-140, FR-PHENO-2015-014, IPPP-15-70, MCNET-15-23, ZU-TH-41-15, MITP-15-108",
    doi = "10.1007/JHEP04(2016)021",
    journal = "JHEP",
    volume = "04",
    pages = "021",
    year = "2016"
}

@article{Lindert:2022qdd,
    author = "Lindert, Jonas M. and Lombardi, Daniele and Wiesemann, Marius and Zanderighi, Giulia and Zanoli, Silvia",
    title = "{W$^{\pm}$Z production at NNLO QCD and NLO EW matched to parton showers with MiNNLO$_{PS}$}",
    eprint = "2208.12660",
    archivePrefix = "arXiv",
    primaryClass = "hep-ph",
    reportNumber = "MPP-2022-98",
    doi = "10.1007/JHEP11(2022)036",
    journal = "JHEP",
    volume = "11",
    pages = "036",
    year = "2022"
}

@article{Lindert:2023fcu,
    author = "Lindert, Jonas M. and Mai, Lorenzo",
    title = "{Logarithmic EW corrections at one-loop}",
    eprint = "2312.07927",
    archivePrefix = "arXiv",
    primaryClass = "hep-ph",
    doi = "10.1140/epjc/s10052-024-13430-0",
    journal = "Eur. Phys. J. C",
    volume = "84",
    number = "10",
    pages = "1084",
    year = "2024"
}

@article{Lindert:2026ief,
    author = "Lindert, J. M. and Mai, L.",
    title = "{Logarithmic EW corrections at two-loop}",
    eprint = "2604.14320",
    archivePrefix = "arXiv",
    primaryClass = "hep-ph",
    doi = "10.1140/epjc/s10052-026-16141-w",
    journal = "Eur. Phys. J. C",
    volume = "86",
    number = "7",
    pages = "905",
    year = "2026"
}

@article{Manohar:2016nzj,
    author = "Manohar, Aneesh and Nason, Paolo and Salam, Gavin P. and Zanderighi, Giulia",
    title = "{How bright is the proton? A precise determination of the photon parton distribution function}",
    eprint = "1607.04266",
    archivePrefix = "arXiv",
    primaryClass = "hep-ph",
    reportNumber = "CERN-TH-2016-155",
    doi = "10.1103/PhysRevLett.117.242002",
    journal = "Phys. Rev. Lett.",
    volume = "117",
    number = "24",
    pages = "242002",
    year = "2016"
}

@article{Muck:2016pko,
    author = {M{\"u}ck, Alexander and Oymanns, Lennart},
    title = "{Resonace-improved parton-shower matching for the Drell-Yan process including electroweak corrections}",
    eprint = "1612.04292",
    archivePrefix = "arXiv",
    primaryClass = "hep-ph",
    reportNumber = "TTK-16-48",
    doi = "10.1007/JHEP05(2017)090",
    journal = "JHEP",
    volume = "05",
    pages = "090",
    year = "2017"
}

@article{Nason:2004rx,
    author = "Nason, Paolo",
    title = "{A New method for combining NLO QCD with shower Monte Carlo algorithms}",
    eprint = "hep-ph/0409146",
    archivePrefix = "arXiv",
    reportNumber = "BICOCCA-FT-04-11",
    doi = "10.1088/1126-6708/2004/11/040",
    journal = "JHEP",
    volume = "11",
    pages = "040",
    year = "2004"
}

@article{Pagani:2021iwa,
    author = "Pagani, Davide and Shao, Hua-Sheng and Tsinikos, Ioannis and Zaro, Marco",
    title = "{Automated EW corrections with isolated photons: t$ \overline{t} ${\ensuremath{\gamma}}, t$ \overline{t} ${\ensuremath{\gamma}}{\ensuremath{\gamma}} and t{\ensuremath{\gamma}}j as case studies}",
    eprint = "2106.02059",
    archivePrefix = "arXiv",
    primaryClass = "hep-ph",
    reportNumber = "DESY 21-083, DESY-21-083, LU-TP 21-19, TIF-UNIMI-2021-7",
    doi = "10.1007/JHEP09(2021)155",
    journal = "JHEP",
    volume = "09",
    pages = "155",
    year = "2021"
}

@article{Pagani:2021vyk,
    author = "Pagani, Davide and Zaro, Marco",
    title = "{One-loop electroweak Sudakov logarithms: a revisitation and automation}",
    eprint = "2110.03714",
    archivePrefix = "arXiv",
    primaryClass = "hep-ph",
    reportNumber = "DESY-21-145, TIF-UNIMI-2021-13",
    doi = "10.1007/JHEP02(2022)161",
    journal = "JHEP",
    volume = "02",
    pages = "161",
    year = "2022"
}

@article{Pagani:2023wgc,
    author = "Pagani, Davide and Vitos, Timea and Zaro, Marco",
    title = "{Improving NLO QCD event generators with high-energy EW corrections}",
    eprint = "2309.00452",
    archivePrefix = "arXiv",
    primaryClass = "hep-ph",
    reportNumber = "TIF-UNIMI-2023-25, LU-TP-23-10",
    doi = "10.1140/epjc/s10052-024-12836-0",
    journal = "Eur. Phys. J. C",
    volume = "84",
    number = "5",
    pages = "514",
    year = "2024"
}

@article{Platzer:2011bc,
    author = "Platzer, Simon and Gieseke, Stefan",
    title = "{Dipole Showers and Automated NLO Matching in Herwig++}",
    eprint = "1109.6256",
    archivePrefix = "arXiv",
    primaryClass = "hep-ph",
    reportNumber = "DESY-11-162, KA-TP-24-2011, HERWIG-11-01, MCNET-11-24",
    doi = "10.1140/epjc/s10052-012-2187-7",
    journal = "Eur. Phys. J. C",
    volume = "72",
    pages = "2187",
    year = "2012"
}

@article{Bothmann:2021led,
    author = "Bothmann, Enrico and Napoletano, Davide and Sch{\"o}nherr, Marek and Schumann, Steffen and Villani, Simon Luca",
    title = "{Higher-order EW corrections in ZZ and ZZj production at the LHC}",
    eprint = "2111.13453",
    archivePrefix = "arXiv",
    primaryClass = "hep-ph",
    doi = "10.1007/JHEP06(2022)064",
    journal = "JHEP",
    volume = "06",
    pages = "064",
    year = "2022"
}

@article{Sherpa:2024mfk,
    author = "Bothmann, Enrico and others",
    collaboration = "Sherpa",
    title = "{Event generation with Sherpa 3}",
    eprint = "2410.22148",
    archivePrefix = "arXiv",
    primaryClass = "hep-ph",
    reportNumber = "IPPP/24/67, LTH-1385, FERMILAB-PUB-24-0748-T, ZU-TH 51/24, MCNET-24-17, CERN-TH-2024-171",
    doi = "10.1007/JHEP12(2024)156",
    journal = "JHEP",
    volume = "12",
    pages = "156",
    year = "2024"
}

\end{document}